\pdfoutput=1
\documentclass[journal]{IEEEtran}

\ifCLASSINFOpdf
\else
\fi

\let\chapter\section
\usepackage{amsmath}
\usepackage{graphicx}
\usepackage{subfigure}
\usepackage{multirow}
\usepackage{amssymb}
\usepackage{algorithmic}
\usepackage{caption}
\usepackage{color}
\usepackage[ruled,lined]{algorithm2e}
\usepackage{bm}
\usepackage{enumerate}
\usepackage{mathrsfs}
\usepackage{leftidx}
\usepackage{authblk}
\usepackage{booktabs}
\usepackage{threeparttable}
\usepackage{array}
\usepackage{amsfonts,amssymb}
\usepackage{hyperref}
\usepackage{textcomp}
\usepackage{xr}
\usepackage{cite}

\usepackage{tikz}
\usetikzlibrary{positioning}
\begin{document}

\captionsetup[table]{name={TABLE},labelsep=none}
\captionsetup[figure]{name={Fig.},labelsep=period}

\title{A Novel Rapid-flooding Approach with Real-time Delay Compensation for Wireless Sensor Network Time Synchronization}
\author{
       Fanrong~Shi,~\IEEEmembership{Member,~IEEE},
       Simon~X.~Yang,~\IEEEmembership{Senior Member,~IEEE},
       Xianguo~Tuo,
       Lili~Ran,
       Yuqing~Huang

\thanks{Manuscript received Aug. 24, 2019; revised Mar. 16, 2020; accepted Apr. 8, 2020. This work was supported in part by National Natural Science Foundation of China Programs (61601383), in part by Key R$\&$D projects of Sichuan Science and Technology Program (No. 20ZDYF2341), in part by Natural Science Foundation of Southwest University of Science and Technology (SWUST) (19zx7157) and Longshan academic talent research supporting program of SWUST (17LZX650, 18LZX633). (Corresponding author: Simon X. Yang.)
}

\thanks{Fanrong Shi, Lili Ran, and Yuqing Huang are with Robot Technology Used for Special Environment Key Laboratory of Sichuan Province, School of Information Engineering, SWUST, Mianyang 621010, China (e-mail: sfr\_swust@swust.edu.cn, lili\_ran@126.com, hyq-851@163.com).}

\thanks{Simon X. Yang is with Advanced Robotics and Intelligent Systems (ARIS) Laboratory, School of Engineering, University of Guelph, Guelph, Ontario, N1G 2W1, Canada (email: syang@uoguelph.ca). }

\thanks{Xianguo Tuo is with Sichuan University of Science and Engineering, Zigong 643000, China (e-mail:tuoxg@cdut.edu.cn).}
}

\markboth{DOI: 10.1109/TCYB.2020.2987758}
{Shell \MakeLowercase{\textit{et al.}}: Bare Demo of IEEEtran.cls for IEEE Journals}
\maketitle

\begin{abstract}
One-way-broadcast based flooding time synchronization algorithms are commonly used in wireless sensor networks (WSNs). However, the packet delay and clock drift pose challenges to accuracy, as they entail serious by-hop error accumulation problems in the WSNs. To overcome it, a rapid-flooding multi-broadcast time synchronization with real-time delay compensation (RDC-RMTS) is proposed in this paper. By using a rapid-flooding protocol, flooding latency of the referenced time information is significantly reduced in the RDC-RMTS. In addition, a new joint clock skew-offset maximum likelihood estimation is developed to obtain the accurate clock parameter estimations, and the real-time packet delay estimation. Moreover, an innovative implementation of the RDC-RMTS is designed with an adaptive clock offset estimation. The experimental results indicate that, the RDC-RMTS can easily reduce the variable delay and significantly slow the growth of by-hop error accumulation. Thus, the proposed RDC-RMTS can achieve accurate time synchronization in large-scale complex WSNs.

\end{abstract}


\IEEEpeerreviewmaketitle

\section{Introduction}
\IEEEPARstart{S}{ynchronization} is the typical problem in distributed system, and it still remains a popular topic in smart grid \cite{PDU_Grid}, Internet of Things (IoT) \cite{IoT1,IoT2}, wireless sensor networks (WSNs) \cite{WSN}, networked synchronization control \cite{DisControl2,DisControl4}, and so on. Specifically, the time synchronization, which aims to correct the local time information on the distributed nodes and drive a common time notion network-wide, is an essential portion of some WSN/IoT applications, e.g., providing accurate time interval for distributed data acquisition and processing \cite{DataSync,Sleep1}, and scheduling \cite{Scheduling}; measuring time delay in industrial wireless feedback control system \cite{SyncCtrl}. However, the large-scale wireless network application imposes exceedingly strict requirements on time synchronization, for instance, they require an accuracy of about $\pm100$ {\textmu s} and $\pm96$ {\textmu s} in the industrial automation standards ISA100.11a \cite{ISA-100.11a} and WirelessHART \cite{WirelessHART}, respectively; about 120 {\textmu s} for the damage localization in the structural health monitoring \cite{SHM_erro,SHM_S}. Considering WSN is being used in many applications, e.g., industrial automation, monitoring, and is resource-constrained, low-cost, large scale and unreliable connection, therefore its time synchronization problem is representative.

The early researches focused on developing novel synchronization structures, reducing the dependence of the algorithms on the network topology management, and improving the robustness of the algorithms in dynamic networks.
As is known to all, due to the requirement of additional protocol to manage the topology of network, the RBS \cite{RBS} and the TPSN \cite{TPSN} are difficult to be deployed in a large-scale complex network, which has large diameter and dynamic topology.
Many of the derived algorithms were expected to optimize the implementation complexity, e.g., the clustered-based algorithm \cite{cluster-Conse-sync-1,PulseSS,cluster-Conse-sync-2,cluster-Conse-sync-3} and low-power interested approach in \cite{CESP}, the improved two-way message exchange time synchronization approaches in \cite{TSync,Timing-sync,Star-Structure,Elsharief2017DensityTable,R-Sync,Q_TTME,TT_K,TT_W}.

In recent years, the flooding time synchronization algorithms \cite{FTSP, FCSA, PulseSync, RMTS} and the consensus-based approaches \cite{ATS2007A,ATS,DCTS,GTSP} are widely studied to meet the accurate synchronization in the large-scale complex WSN. These time synchronization algorithms do not require any topology management, so the problems mentioned above are naturally avoided. Moreover, they are robust and scalable, and can adapt quickly to the changes in network connectivity and clock drifts. However, there are limitations to both of the flooding and the consensus-based methods. In other words, they are different from each other and can be used to meet the time synchronization requirements in different WSNs.

Considering a consensus-based time synchronization algorithm, it is easy to implement and robust in dynamic networks. However, its convergence rate is very slow due to the iteration. For instance, the convergence time is up to 120 rounds of synchronization intervals in ATS ($5\times7$ grid network) \cite{ATS}, and studies in \cite{CCS,EventDriven_ATS,MTS,DiStiNCT,HE2017C,MACTS} are interested in improving the convergence rate. However, It cannot to be synchronized by an external clock or the heterogeneous network \cite{EGsync}. What it matters is not the value of the reference clock but the fact that all clocks converge to one common virtual reference, thus it cannot be synchronized with an external clock or the heterogeneous network \cite{EGsync}.

Fortunately, flooding time synchronization algorithm can easily avoid the problems in a consensus-based method. It employs a special node as the reference, meanwhile it distributes the reference's time information network-wide and aims to synchronize all of the nodes to the reference. A number of line networks are automatically generated while the referenced time information is flooded. Therefore, a complex network is simplified as lines, and the distance of line is the main factor of network to affect the algorithm. The rapid-flooding protocol leads the algorithms to build a stable synchronization after a few synchronization intervals, e.g., about 3 rounds in RMTS ($5\times5$ grid network) \cite{RMTS}. Moreover, the rapid-flooding is a synchronous method in which the time interval of node transmission can be predicted, thus it is possible and easy to meet synchronous sleep/wake scheduling for low-duty cycle WSN \cite{Chase,Splash}. Hence, the flooding time synchronization algorithm will be a very efficient solution for the time synchronization in the large-scale WSNs.

However, distance of flooding path, flooding latency, packet delay, and clock drift, which cause the unpredictable by-hop error accumulation problem, pose the main challenges to a flooding time synchronization algorithm. Such problems have been significantly improved by the previous studies, and most of the recent studies on flooding time synchronization focus on improving the accuracy, e.g., PulseSync \cite{PulseSync_1,PulseSync}, Glossy \cite{glossy}, FCSA \cite{FCSA}, RMTS \cite{RMTS}. The improvements of flooding time synchronization are simple and effective, i.e., shortening the flooding latency and removing the delay, meanwhile compensating the clock drift. However, there is no better way to calculate packet delay in existing one-way broadcast-based synchronization protocols, but using the prior constant, e.g., the method used in PulseSync and RMTS. Hence, before deploying WSN, a lot of preliminary experiments are required to calculate the prior constant; actual delay may change due to hardware and environment changes.

In this paper, we are interested in a real-time and agile delay estimation to improve the rapid-flooding time synchronization. If the clock skew estimation is fast convergence and independent of the clock offset estimation, meanwhile if a two-way message exchange can be constructed in the one-way-broadcast model, then the real-time delay compensation can be implemented. It is worth mentioning that this strategy has been inspired by the two-way message exchange model based clock offset estimation in \cite{DRJeskeMLE,KLNohMLE,QMChaudhariMLE}. To achieve the objective, the Maximum Likelihood Estimation (MLE) proposed in \cite{MLE} is employed to generate an accurate clock skew estimation at the second round of synchronization; meanwhile an improved two-way message exchange model is constructed on one-way-broadcast model; then a joint clock skew-offset MLE is obtained , and the real-time delay estimation is also given. As a result, a new rapid-flooding multi-broadcast time synchronization with real-time delay compensation (RDC-RMTS) protocol is further developed. The main contributions of this work are as follows.

\emph{1)} A novel two-way message exchange model is ingeniously designed in the one-way broadcast-based flooding time synchronization, which utilizes redundant broadcasting without any additional packet transmission.

\emph{2)} The proposed joint clock skew-offset MLE provides the real-time delay estimations and the accurate clock offset estimation, which results in better scalability to RDC-RMTS.

\emph{3)} An innovate implementation is developed for the RDC-RMTS, in which an adaptive clock offset estimation is designed to guarantee the accurate estimation over unreliable network.

Moreover, the actual performances of flooding time synchronization in a large-scale complex network are discussed. Basically, the performance evaluation indicate that the network mainly affect the time synchronization in a way of increasing the flooding path.

The rest of the paper is organized as follows. In Section II, the challenges to the flooding time synchronization is discussed in detail, and the motivation of this paper is provided. The system model is provided in Section III and the RDC-RMTS is proposed in Section IV. The implementation of RDC-RMTS is described in Section V. The RDC-RMTS is evaluated and discussed in Section VI, where the testbed is provided, and the experimental results in FTSP, FCSA, FloodPISync, PulseSync, PulsePISync, and RMTS are also reported. The discussions and simulations results in complex network are reported in Section VII. Finally, conclusions are drawn in Section VIII.

\section{Challenges and Motivation}

Flooding time synchronization protocol is easy to implement, and does not require topology management. In addition, flooding operations decompose a complex network into multi-line networks, then the nodes on the line synchronize itself to the reference node (root). However, the time synchronization on a multi-hop node depends on the time information forwarded by relay nodes, thus relay node synchronization error will accumulate on a hop-by-hop basis, i.e., by-hop error accumulation problem \cite{RMTS}, which is the major challenge to flooding time synchronization.

A by-hop error accumulation is mainly caused by flooding latency, packet transmission delay, and clock drift. In addition, it is proportional to the distance of reference node and multi-hop node. Definition $E$ is the synchronization error of the node relative to the reference node, then the synchronization error on the $k$-hop node $E[k]$ which is derived in \cite{RMTS}, is given by
\begin{equation}\label{equ:1}
 E[k]=\sum_{h=1}^{k}D[h]+\frac{\sum_{h=1}^k \varphi_h^{h-1}\times T_{wait}[h]}{10^6}
\end{equation}
where $T_{wait}[h]$, which is less than the synchronization interval ($period$), is the flooding latency at the $h$-hop node, and $\varphi_h^{h-1}$ is the relative clock rate between $h$-hop and ($h-1$)-hop.Variable $D[h]$ is one-way broadcast packet transmission delay at the $h-$hop node, which comprises variable portion $D_{var}$ and uncertain portion $D_{unc}$ \cite{RMTS}. Moreover, $D_{var}=D_{fixed}+d$, where constant $D_{fixed}$ is the mean of $D_{var}$, and variable $d\sim N(0,\sigma^2)$ ($\sigma$ is the standard deviation of $D_{var}$). Hence, $D=D_{fixed}+d+D_{unc}$. Because relative clock rate or clock frequency offset is usually described in parts per million (PPM), a multiplier $10^6$ is used in (\ref{equ:1}).

Summaries of the existing flooding time synchronization algorithms are shown in Table \ref{tab:1}. It should be noted that, according to the \cite{FCSA, PISync}, although they don't seems emphasize the use of $\hat{D}_{fixed}$ to compensate clock offset estimation, we prefer to do this in our experiments to get better experimental results. Moreover, $\hat{D}_{fixed}$ is calculated during the experimental phase based on the previous experimental results, and it is a constant when the algorithm is evaluated.

\vspace{0.8cm}
\begin{table}[htbp]\scriptsize
 \centering
 \captionsetup{justification=centering}
 \caption{\\Summary of the flooding time synchronization algorithms in comparison. Considering the pairwise timestamps $T_s$ (created at sender) and $T_r$ (created at receiver), clock offset estimations $\hat{\theta}_1=T_s-T_r=\theta+D$, $\hat{\theta}_2=\hat{\theta}_1-\hat{D}_{fixed}$, and $\hat{\theta}_3=\min\{\hat{\theta}_1[n]\}_{n=1}^N-\hat{D}_{fixed}$.}{\label{tab:1}}
 {
 \begin{tabular}{ccccccccc}
  \toprule
  \toprule
               & \multirow{2}{*} { \textbf{Protocol }}      &  \multicolumn{2}{c}{ \textbf{Clock Parameter Estimations }} \\
               &                                    &{ Clock Offset}                & Clock Skew \\
  \midrule
    FTSP	       &Slow-flooding                        &$\hat{\theta}_1$	 	                       & LR       \\
    FCSA	       &Slow-flooding 	                     &$\hat{\theta}_2$	  	                       & LR       \\
    FloodPISync    &Slow-flooding 	                     &$\hat{\theta}_2$	                           & PI       \\
    PulseSync      &Rapid-flooding                       &$\hat{\theta}_2$	   	                       & LR       \\
    PulsePISync    &Rapid-flooding                       &$\hat{\theta}_2$	  	                       & PI       \\
    RMTS           &Rapid-flooding                       &$\hat{\theta}_3$ 	  	                       & MLE      \\
  \bottomrule
  \bottomrule
 \end{tabular}
 }
\end{table}

Lenzen et al. pointed out that, when the referenced time information is flooded network-wide, it will lose accuracy due to clock drift and flooding latency \cite{PulseSync_1}. The PulseSync suggests to distribute time information as fast as possible, and is expected to adapt to fast change in clock drift. Unlike FTSP, which uses slow-flooding, PulseSync uses a rapid-flooding protocol. The Glossy and RMTS are also rapid-flooding approaches.

The FCSA tries to minimize the undesired effect of flood waiting times on the synchronization accuracy by using a clock rate agreement algorithm, and expects to force all nodes to agree on a common clock rate. Even so, the slight estimation error on clock skew estimation may cause serious interference to time synchronization accuracy. Especially, the FCSA use linear regression (LR) to estimate the clock skew, where the $D$ is directly introduced in observations. By using LR, the problem in FCSA also exists in FTSP and PulseSync.
The PISync employs a Proportion Integration (PI) control method to optimize the clock skew estimation, however the delays in the observations are not optimized. Unlike that, the MLE in \cite{MLE,RMTS} tries to minimize the delay and could obtain more accurate clock skew estimation than LR and PI method \cite{MLE}.

The existing flooding time synchronization algorithms use accurate clock parameter estimation methods and rapid-flooding protocol, and are expected to against the by-hop error accumulation problem. Considering the by-hop error accumulation model in (\ref{equ:1}), the FCSA tries to minimize $\varphi_h^{h-1}$, moreover, the PulseSync and RMTS are also to minimize $T_{wait}$. As discussed in \cite{RMTS}, the summary of the synchronization error in flooding time synchronization algorithms is shown in \ref{tab:2}. It is clear that, the accuracy is significantly improved in the existing flooding time synchronization algorithms.

\vspace{0.8cm}
\begin{table}[htbp]\scriptsize
 \centering
 \captionsetup{justification=centering}
 \caption{\\Summary of the synchronization error in flooding time synchronization algorithms, where $D=D_{fixed}+d+D_{unc}$. The synchronization error in FloodPISync and PulsePISync are similar to FCSA and PulseSync, respectively.}{\label{tab:2}}
 {
 \begin{tabular}{ccccccccc}
  \toprule
  \toprule
                 & \multicolumn{2}{c}{\textbf{Synchronization Error }}              \\
                &{ Single-hop }      & $k$-hop                                      \\
  \midrule
    FTSP	   	   & $D$                &  (\ref{equ:1})                           \\
    FCSA	       & $D$               &  $\approx\sum_{h=1}^{k}D[h]$               \\
    PulseSync      & $D-D_{fixed}$     &  $\approx\sum_{h=1}^{k}(D[h]-D_{fixed})$   \\
    RMTS           & $d$               &  $\approx\sum_{h=1}^{k}d[h]$               \\
  \bottomrule
  \bottomrule
 \end{tabular}
 }
\end{table}

PulseSync and RMTS use a fixed delay estimation for clock offset estimate compensation that pose challenges to flexible deployment, as it entails that a lot of preliminary experiments need to be completed for the delay estimation. Testbed is needed to collect experimental results and calculate delay estimation, and the estimation cannot be changed after being deployed. Thus, the synchronization accuracy of algorithm may decrease results from the timely changes in delay.

In the light of the above discussion, the main motivation of this paper is to obtain a timely delay estimation, and then propose a new rapid-flooding time synchronization algorithm with real-time delay compensation against the by-hop error accumulation problem.

\section{System Model}

The WSN is modeled as the graph $\mathcal{G}=(\mathcal{N},\mathcal{E})$, where $\mathcal{N}=\{1,2,\ldots,n\}$ represents the nodes of the WSN and $\mathcal{E}$ defines the available communications links. The set of neighbors for $v_i$ is $\mathcal{N}_i=\{j|(i,j)\in \mathcal{E},i\neq j\}$, where nodes $v_i$ and $v_j$ belong to $\mathcal{N}$, and $j\in \mathcal{N}_i$. In our work, there are two time notions defined for the time synchronization algorithm, i.e. the hardware clock notion $H(t)$ and logical clock notion $L(t)$. Hardware clock $H(t)$ is defined as
\begin{equation}\label{equ:2}
  H(t)=\int_0^t h(\tau)d\tau+\theta(t_0)
\end{equation}
where $H(t)$ is the hardware clock notion. Parameter $h(\tau)$ is the hardware frequency rate (clock speed) of the clock source, meanwhile it is the inherent attribute of crystal oscillator and cannot be changed or measured. Any node considers itself to have the ideal clock frequency (i.e. nominal frequency) and $h(\tau)=1$. Variable $t_0$ is the moment that node is powered on, and $\theta(t_0)$ is the initial relative clock offset. It should be noticed that $H(t)$ cannot be changed either, and the timestamps on $H(t)$ are used to estimate the relative clock speed for the proposed algorithm.

The logical clock $L(t)$ is defined as
\begin{equation}\label{equ:3}
  L(t)=\varphi(t)\times H(t)
\end{equation}
where $\varphi(t)$ is the logical relative clock rate and initialized as 1, it can be changed to speed up or slow down $L(t)$. With respect to the reference node, $\varphi(t)$ is always set as 1. The timestamps on $L(t)$ are used to estimate the relative clock offset for the proposed algorithm. Variable $L(t)$ is used to supply the global time service for the synchronization applications.

Considering the arbitrary nodes $v_i$ and $v_j$, $L_i(t)$ and $L_j(t)$ are the logical times respectively, and $\varphi_i^j$ is the relative logical frequency rate which is\begin{equation}\label{equ:4}
  \varphi_i^j=1+\frac{\theta_\Delta}{t_\Delta},  t_\Delta>0
\end{equation}
where, in consideration of arbitrary moments $t_1$ and $t_2(t_1<t_2)$, $t_\Delta=t_2-t_1$. The relative clock offset increment in $t_\Delta$ is $\theta_\Delta=(H_j(t_2 )-H_i(t_2 ))-(H_j(t_1 )-H_i(t_1 ))$. Parameter $\varphi_i^j$ is the changing rate on the relative clock offset in $t_\Delta$. It is common estimated by any two clock offset estimations and used to compensate the clock skew in the time synchronization algorithms.

\section{Proposed RDC-RMTS Algorithm}
Considering the one-way broadcast based time synchronization protocols, and assuming that the clock skew is fixed in a short time, the key idea of RDC-RMTS is that: the clock skew estimation is independent of clock offset estimation, and convergence fast and accurate, then; an improved two-way message exchange can be structured for the clock offset estimation based on the one-way broadcasts, and the delay can be estimated and compensated timely in the RDC-RMTS.

\subsection{Proposed Synchronization Model of RDC-RMTS}

\begin{figure}[!htb]
\centering
\includegraphics[scale=0.5]{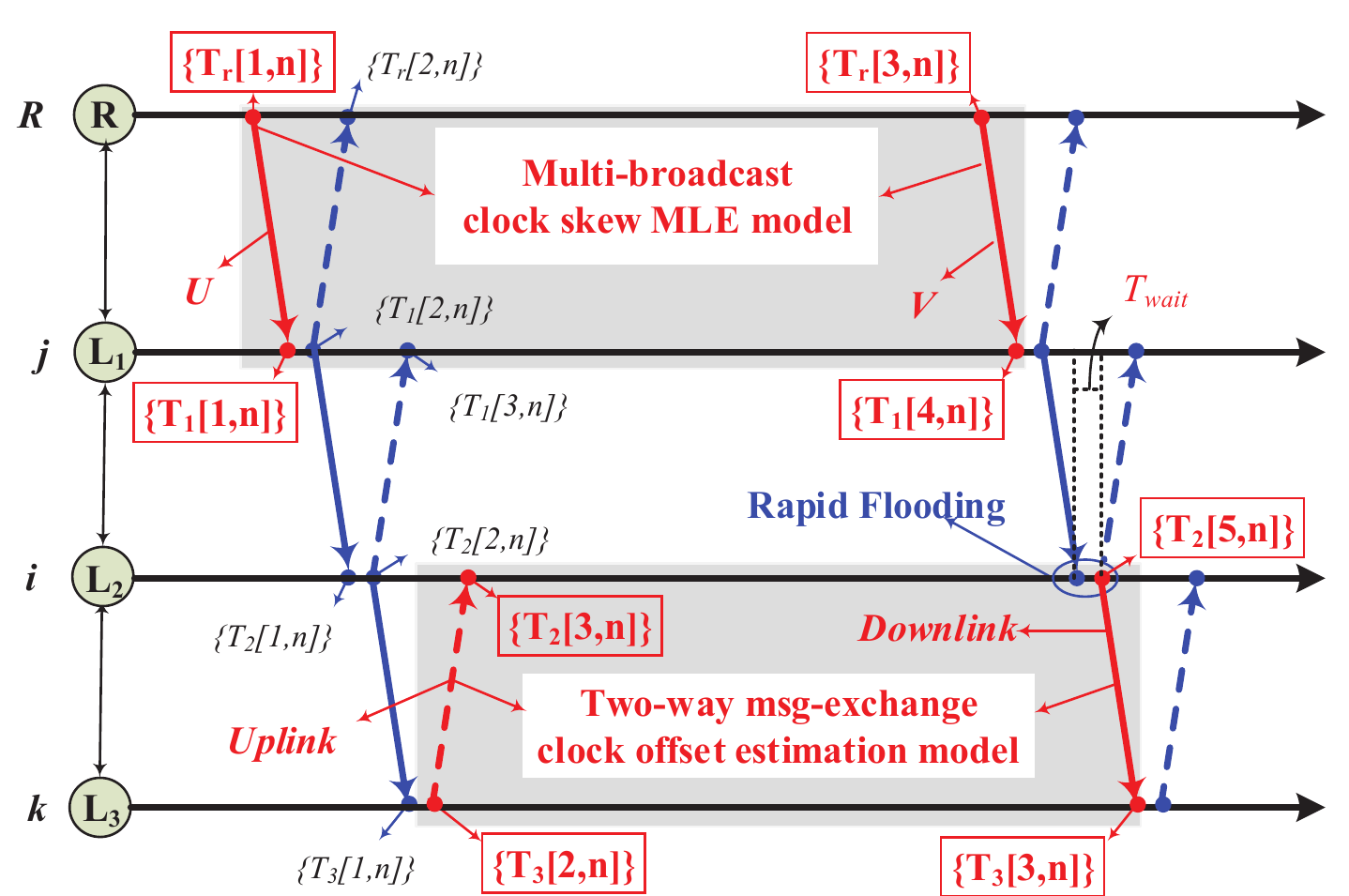}
\caption{The multi-one-way broadcast model. Nodes broadcast $n$ time information packets in a short time, and n pairs of timestamps are generated, e.g. $T_r[1,1]$ and $T_1[1,1]$ at nodes $v_R$ and $v_j$ respectively.}
\label{SynchModel}
\end{figure}

In this part, an rapid-flooding synchronization model is proposed for the RDC-RMTS based on multiple one-way broadcast, which is illustrated in Fig. \ref{SynchModel}. There are two important portions in the proposed model, i.e. multi-broadcast clock skew estimation and two-way message exchange clock offset estimation. The main difference between the new proposed synchronization model and the previous ones is that a novel two-way message exchange model is structured in the one-way broadcast synchronization model.

\vspace{0.3cm}

Considering the arbitrary $v_i\in \mathcal{G}$ periodically broadcasts a group of $N$ synchronization packets to neighboring nodes over a broadcast period of $T_b$. Here we define $\{T_k[x,n]\}_{n=1}^N$ as the set of timestamps at node $k(k\in \mathcal{G})$, its size is $N$, and $x$ is its serial number. There are $N$ packets broadcasted in a short time, and $N$ pairs of timestamps are generated. When a node broadcasts a packets, the timestamp will be created at both of the sender and the receivers.

Clock skew MLE of the RDC-RMTS is developed based on the  multi-broadcast clock skew estimation, and uses two groups of multi-broadcast process (e.g., $U$ and $V$) to collect observations, e.g. $\langle\{T_r[1,n]\}_{n=1}^N,\{T_1[1,n]\}_{n=1}^N \rangle$ and $\langle\{T_r[2,n]\}_{n=1}^N,\{T_1[2,n]\}_{n=1}^N \rangle$. Clock offset estimation of RDC-RMTS is developed based on the two-way message exchange model and the above clock skew estimation, e.g.$\langle\{T_3[2,n],T_2[3,n]\}_{n=1}^N$ and $\langle\{T_2[5,n],T_3[3,n]\}_{n=1}^N$. Since that the time interval from the first broadcast to latest broadcast (e.g., $T_r[1,N]-T_r[1,1]$, when $N=5$, which is less than 12 microsecond in our experiments) is exceedingly short, we can assume that the relative clock offset is fixed at the phase of a multi-broadcast processing.

The RDC-RMTS is a rapid flooding time synchronization approach: the reference node $v_R$ initializes an synchronization period and nodes forward the newly time information to its neighbors quickly (the message flooding latency $T_{wait}$ is very short).

\subsection{The Clock Skew Estimation}
Assuming the actual delay is Gaussian distribution, the clock skew MLE is proposed in \cite{MLE}.

In this paper, the proposed synchronization model in Fig. \ref{SynchModel} provides an reasonable implementation for the MLE, i.e. the multi-broadcast clock skew estimation model (synchronization process $U$ and $V$). A group of timestamp observations is created in the multi-broadcast process, i.e., $\langle\{T_r[1,n],T_1[1,n]\}_{n=1}^N \rangle$ and $\langle\{T_r[3,n],T_1[3,n]\}_{n=1}^N\rangle$, then the observations $P$ for the clock skew MLE likelihood function are
\begin{equation}\label{equ:5}
  P:p[n]=v[n]-u[n], n=1,2,\ldots,N
\end{equation}
where $u[n]=T_1[1,n]-T_r[1,n]$, $v[n]=T_1[3,n]-T_r[3,n]$.

Accordingly, the clock skew MLE $\hat{\varphi}_{i(MLE)}^r$ is
\begin{equation}\label{equ:6}
\hat{\varphi}_{i(MLE)}^r=\frac{\bar{P}}{\hat{\tau}}=\frac{\sum_{n=1}^N p[n]}{N\hat{\tau}}
\end{equation}
where $\tau$ is the time interval of $U$ and $V$, e.g., $\hat{\tau}=T_1[3,1]-T_1[1,1]$.

It is worth mentioning that the $\hat{\varphi}_{i(MLE)}^r$ is independent of clock offset estimation, and the timestamps for $\hat{\varphi}_{i(MLE)}^r$ is generated at the hardware clock timer.

\subsection{Proposed Joint Clock Skew-offset MLE}
According to \cite{TPSN}, the clock offset estimation based on two-way message exchange model significantly optimizes the estimation error caused by delay, and more accurate than that of one-way broadcast model. Most of the previous clock offset MLEs, e.g., \cite{DRJeskeMLE,KLNohMLE,QMChaudhariMLE}, are developed based on two-way message exchange model. However, the traditional two-way message exchange model is implemented on a pair of nodes, thus the reference should synchronize the neighbors one by one, and it may fail to meet the efficient time synchronization over a large-scale network due to the pairwise message exchange and hierarchical management.

In this part, we employ the MLE clock skew to calculate the clock offset estimation, which is significantly different from the traditional method, e.g. the FTSP in which the linear regression is used to calculate clock skew estimations. It is worthy to note that, the MLE clock skew is independent of the clock offset estimation, therefore it is possible to use the accurate clock skew estimation to compensate the clock offset estimation.

According to the two-way message exchange clock offset estimation model that is shown in Fig. \ref{SynchModel}, for $v_i$ and $v_k$, the $Uplink$ timestamps are $\langle\{T_3[2,n],T_2[3,n]\}_{n=1}^N\rangle$, and the $Downlink$ timestamps are $\langle\{T_2[5,n],T_3[3,n]\}_{n=1}^N\rangle$. We define that $\{p_u[n]\}_{n=1}^N$ and $\{p_d[n]\}_{n=1}^N$ as
\begin{equation}\label{equ:7}
    p_u[n]=T_2[3,n]-T_3[2,n]=D_u[n]+\theta_\Delta-\theta_d,
\end{equation}
\begin{equation}\label{equ:8}
    p_d[n]=T_3[3,n]-T_2[5,n]=D_d[n]+\theta_d
\end{equation}
where $\theta_d$ is the relative clock offset in $downlink$, and $\theta_\Delta=\varphi_k^i\times T_b$, then the two-way message exchange clock offset estimation can be rewritten as
\begin{equation}\label{equ:9}
    \hat{\theta}_d[n]= \frac{(p_d[n]-p_u[n])-(D_d[n]-D_u[n])+\theta_\Delta}{2},
\end{equation}
\begin{equation}\label{equ:10}
    \hat{D}_d[n]= \frac{(p_d[n]+p_u[n])-\theta_\Delta}{2}.
\end{equation}

Considering that $D$ is variable value with fixed portion $D_{fixed}$ and variable portion $d$, and using the $\hat{\varphi}_{k(MLE)}^i$ (i.e., $\hat{\theta}_\Delta=\hat{\varphi}_{k(MLE)}^i\times \tau$, the clock offset estimation MLE of clock offset $\hat{\theta}_{k}^i$ is
\begin{equation}\label{equ:11}
    \hat{\theta}_{k}^i=\hat{\theta}_d[n]=\frac{(p_d[n]-p_u[n])-(d_d[n]-d_u[n])+\hat{\theta}_\Delta}{2}.
\end{equation}

According to \cite{DRJeskeMLE}, the MLE of joint clock skew-offset estimation is
\begin{equation}\label{equ:12}
    \hat{\theta}_{k(MLE)}^i= \frac{\min \{p_d[n]\}_{n=1}^N - \min \{p_u[n]\}_{n=1}^N+\hat{\theta}_\Delta}{2}
\end{equation}
where the $D_{fixed}$ is removed from the clock offset estimation. The $\hat{D}_{fixed}$ is given by
\begin{equation}\label{equ:13}
    \hat{D}_{fixed}= \frac{\min \{p_d[n]\}_{n=1}^N+ \min \{p_u[n]\}_{n=1}^N-\hat{\theta}_\Delta}{2}.
\end{equation}

It is worth mentioning that the timestamps for $\hat{\theta}_{k(MLE)}^i$ are generated at the logical clock timer.

\subsection{Min function-based MLE}
According to \cite{RMTS}, it is because $D$ in the observations $u[n]$ and $v[n]$  is always a positive value , i.e., $D>0$, the min function-based MLE tries to find out the observation with minimal value of $D$. For instance, the $\hat{\theta}_{j(MLE)}^r$ given by
\begin{equation}\label{equ:14}
 \hat{\theta}_{j(MLE)}^r=\min_{1\leq n \leq N}{\{u[n]\}_{n=1}^N}-\hat{D}_{fixed}
\end{equation}
where $\hat{D}_{fixed}$ is the delay estimation, which can be calculated by (\ref{equ:13}).

\subsection{Error Analysis}
According to (\ref{equ:12}) and (\ref{equ:14}), the $D_{unc}$ and $D_{fixed}$ could be removed from the error link of the RDC-RMTS. Then the single-hop synchronization error $E_R[1]$ of RDC-RMTS is given by
\begin{equation}\label{equ:15}
 E_R[1]=D-D_{fixed}-D_{unc}\approx d.		
\end{equation}

By using the rapid-flood protocol and accurate clock skew compensation, the part $\frac{\sum_{h=1}^k \varphi_h^{h-1}\times T_{wait}[h]}{10^6}$ in (\ref{equ:1}) is almost equal to 0 ($T_{wait}\rightarrow 0$), thus the by-hop error accumulation $E_R[k]$ in RDC-RMTS is given by
\begin{equation}\label{equ:16}
 E_R[k]\thickapprox \sum_{h=1}^{k}(D[h]-\hat{D}_{fixed}-D_{unc}[h])\thickapprox \sum_{h=1}^{k}d[h]	
\end{equation}
which is the same as that in RMTS \cite{RMTS}.

\section{Implementation}

The proposed RDC-RMTS removes the $D_{fixed}$ from clock offset estimation based on the improved two-way message exchange model without any prior knowledge. It is well known that the two-way message exchange model based synchronization algorithms (e.g. TPSN) calculate the clock offset estimation and delay separately. In this part, we demonstrate an innovative implementation for the two-way message exchange based joint clock skew-offset MLE proposed in (\ref{equ:12}) and (\ref{equ:14}).

\subsection{The Structure Overview of RDC-RMTS}
According to (\ref{equ:6}), it suggests larger observation interval $\tau$ to obtain the more accurate clock skew estimation $\hat{\varphi}$. Thus a sliding window is employed in RDC-RMTS, where $\tau$ is several times the actual synchronization period $T_b$.

\begin{figure}[!htb]
\centering
\includegraphics[scale=0.9]{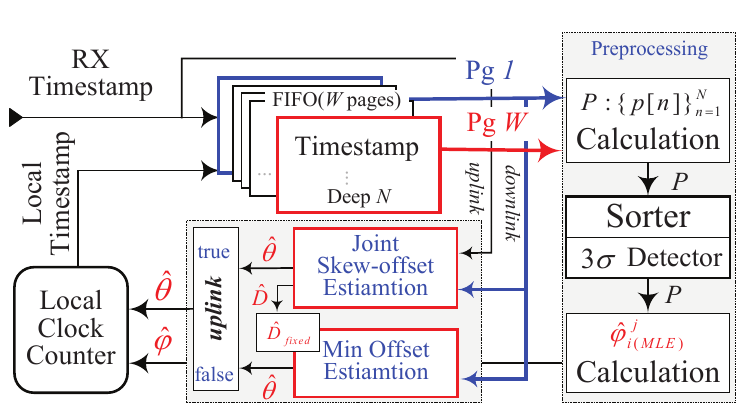}
\caption{The multi-one-way-broadcast model. Nodes broadcast $n$ time information packets in a short time, and n pairs of timestamps are generated, e.g. $T_r(1,1)$ and $T_1(1,1)$ at nodes $v_R$ and $v_j$ respectively.}
\label{MLE_implementation}
\end{figure}

Figure \ref{MLE_implementation} indicates the structure of implementation in RDC-RMTS. Considering $v_i$ and its neighbor $v_j$, it is mainly composed of three parts: sliding window, clock skew estimation, and clock offset estimation.

\vspace{0.3cm}

A $W$-pages ($W\geq2$) timestamp FIFO (first in first out) buffer is the key to the sliding window. It is designed to buffer multiple groups of timestamp, and the sliding window with specified width will be generated for $\hat{\varphi}_{i(MLE)}^j$  by copying the timestamp of the corresponding position of the FIFO. The pages ($W$) of FIFO should be larger than the maximal width of sliding window. The depth of the page is the same as the multiple number $N$. If the sliding window length is set to be $w (w\leq W)$, then the $\hat{\varphi}_{i(MLE)}^j$ is calculated based on the latest timestamp observations (Pg $1$) and the $w$-th from last (i.e.,Pg $w$), and the $\tau$ in (\ref{equ:6}) is $(w-1)$ times of the synchronization period $T_b$, i.e., $\hat{\tau}=(w-1)\times T_b$.

\emph{\textbf{1) The clock skew estimation}}

As discussed above, the $D_{unc}$ is inevitable in the one-way broadcast. Moreover, it is not Gaussian distribution and much larger than the $D_{var}$ (which is Gaussian variable). Thus, the observations need to be preprocessed to meet the assumption (the delay is Gaussian distribution) for $\hat{\varphi}_{i(MLE)}^j$ in (\ref{equ:6}). The $\hat{\varphi}_{i(MLE)}^j$ is implemented in three steps: the observation calculation in (\ref{equ:5}), outlier detector (Sort and $3\sigma$ detector), and $\hat{\varphi}_{i(MLE)}^j$ calculation in (\ref{equ:6}). Since the $D_{unc}$ has extremely low probability, one can assume that the possible $D_{unc}$ samples will be moved to the end of $P$ by the Sort. The $\hat{\sigma}$ can be calculated from the front observations, e.g., $\hat{\sigma}[k]=$std$(\{ p[n]\}_{n=1}^{k-1}), N/2<k\leq N$. Then the possible $D_{unc}$ observations will be recognized and removed from $P$ by outlier detector.

\emph{\textbf{ 2) The adaptive clock offset estimation}}

Considering the large-scale complex WSN, the wireless channel collision probability will increase at $uplink$ in the rapid-flooding. The rapid-flooding protocol requires the higher level nodes to forward the received time information immediately, then it will be a high probability that the adjacent nodes broadcast packets at the same moment. In other words, the low level nodes can not receive all of the $uplink$ messages, and it will cause the $\hat{\theta}_{i(MLE)}^j$ in (\ref{equ:12}) to be unachievable.

To solve this issue, an adaptive approach is developed based on double clock offset estimations, i.e., the joint clock skew-offset MLE, and the min function-based MLE.

\textbf{ Case $\bm{Uplink}$ true:} the RDC-RMTS employs the joint clock skew-offset MLE. The $Downlink$ of two-way message exchange model is the latest multi-broadcast process of $v_j$, and the $Uplink$ is the latest multi-broadcast process of $v_i$, and $\hat{\tau}=T_b$. parameter ${\hat{\theta}_{i(MLE)}^j}$ is calculated by (\ref{equ:12}), and the ${\hat{D}_{fixed)}}$ is calculated by (\ref{equ:13}).

\textbf{ Case $\bm{Uplink}$ false:} the RDC-RMTS employs the min function-based MLE. The observations $\{u[n]\}_{n=1}^N$ of $Downlink$ are used, and ${\hat{\theta}_{i(MLE)}^j}$ is calculated by (\ref{equ:14}), $\hat{D}_{fixed}$ is the average of the valid value calculated by (\ref{equ:13}). The $Downlink$ of two-way message exchange model is the latest multi-broadcast process of $v_j$.

Therefore, when the $Uplink$ works, the $\hat{D}_{fixed}$ is calculated. The RDC-RMTS does not requires any prior experiments to set a fixed $\hat{D}_{fixed}$ just like RMTS, but uses a timely delay compensation. Moreover, the receiver calculates $\hat{D}_{fixed}$ separately for different nodes. Thus, the slight differences on delay caused by the hardware differences of nodes will be handled in RDC-RMTS, while they are directly ignored in PulseSync and RMTS.

\subsection{The Implementation of RDC-RMTS}

In this Section, we discuss the implementation of the proposed RDC-RMTS algorithm in details. For ease of description, considering the pair of adjacent nodes in Fig. \ref{SynchModel}, the node that has the lower level is defined as father, and the higher one as son, e.g. $v_i$ is the father of $v_k$, and $v_j$ is the father of $v_i$. In addition, the sliding window length $W$ is set as 2.

The RDC-RMTS can use the specified node as the root (reference) in WSN applications, e.g., the sink node, or the gateway node (the interface of heterogeneous networks). The flooding time synchronization cannot maintain the network-wide synchronization without an active root. Hence the RDC-RMTS employs a simple root election protocol which is used in \cite{FTSP,FCSA, RMTS} to find a new root when the current root is invalid.

According to the synchronization model (Fig. \ref{SynchModel}) and the implementation structure (Fig. \ref{MLE_implementation}) of RDC-RMTS, a root (reference) algorithm and non-root algorithm are designed for the RDC-RMTS protocol.

\emph{\textbf{1) The root algorithm:}} The periodic synchronization is created at root, and the reference's time information is embedded into the multi-broadcast packets and flooded in the network. Meanwhile, the root will listen for broadcast messages from neighbors and store the timestamps for the joint clock skew-offset estimation.

The pseudo-code of the root algorithm is presented in Algorithm 1, where $\varphi_R$ is the multiplier of $L_R(t)$; parameter $ID_R$ is an identifier of each synchronization process; buffer $\{buf_R[m]\}_{m=1}^M$ is employed to store the corresponding timestamps when root's neighbors forward its time information; the synchronization period is generated at root, i.e., $T_b$.

The $\varphi_R$ is a constant, i.e., $\varphi_R=1$ (Line \ref{alg1.2} in Algorithm 1). While, if an external clock $L_{ER}$ (e.g., GPS ) is used , then the  root can synchronize to $L_{ER}$ (Algorithm 1, Lines \ref{alg1.5} and \ref{alg1.61}), and then its logical time is $L_R(t+\tau)= \varphi_R\times (H_R(t+\tau)-H_R(t))+L_{ER}(t),\tau\geq0$, and change the logical clock rate by adjusting $\varphi_R$.

The root maintains a periodically-scheduled task to initialize the synchronization period and distribute the reference time information packets to neighbors, as Lines \ref{alg1.9}-\ref{alg1.15} in Algorithm 1. The $N$ packets are sent to neighbors rapidly thus the clock offset is almost a fixed value during that interval, as Lines \ref{alg1.9}-\ref{alg1.12} in Algorithm 1. After that, $ID_R$ will be increased by 1 (Line \ref{alg1.14} in Algorithm 1).

Two groups of timestamps are created and embedded to the corresponding packet over the phase of broadcasting, i.e., $H_R[n]$ (created on $H_R(t)$) and $L_R[n]$ (created on $L_R(t)$). The broadcast packets comprise of five parts: $H_R[n]$, $L_R[n]$, $\varphi_R$, $ID_R$, and $buf_R$ (Line \ref{alg1.11} in Algorithm 1).

\LinesNumbered
\begin{algorithm}[ht]
\caption{The root algorithm pseudo-code for RDC-RMTS. $v_R$ is the root, , $j\in \mathcal{N}_R, (R,j)\in \mathcal{G}$. Parameter $N$ is the maximal number of multi-broadcast. Parameter $M$ is maximal numbers of neighbor.}
\if\textbf{Initialization:}                                 {\label{alg1.1}}\\
\quad{Set $\varphi_R=1$; $ID_R=1$;  $n=0$}                  {\label{alg1.2}}\\
\qquad{    $\{buf_R[m]\}_{m=1}^M\leftarrow 0$}               {\label{alg1.3}}\\
\quad{Start periodic broadcast task, period of $T_b$}          {\label{alg1.4}}\\
\BlankLine
\If {(external clock is used)}{                             {\label{alg1.5}}
      {Set $L_R(t)\leftarrow L_{ER}(t)$}\\                {\label{alg1.6}}
    }                                                        {\label{alg1.61}}
\BlankLine
\if\textbf{Upon Receiving $\langle L_j[n], ID_j \rangle$:}                 {\label{alg1.7}}\\
\quad{Store $buf_R(j)\leftarrow \min\limits_{1<n<N}\{L_R[n]-L_j[n]\}$ }           {\label{alg1.8}}\\
\BlankLine
\if\textbf{Upon triggering of broadcast task:}{\\                               {\label{alg1.9}}
  \eIf{($n<N$)} {                                                               {\label{alg1.10}}
         Broadcast $\langle H_R[n], L_R[n],\varphi_R, ID_R, buf_R\rangle $\\    {\label{alg1.11}}
         Set $n=n+1$, go back to (if ($n<N$))                          {\label{alg1.12}}
    }{                                                              {\label{alg1.13}}
         $ID_R= ID_R+1$, $n=0$                                     {\label{alg1.14}}
    }                                                               {\label{alg1.15}}
}
\end{algorithm}

The $\{buf_R[m]\}_{m=1}^M$ is set as 0 at the phase of initialization and the moment after multi-broadcast. When the multi-broadcast is received, it will be updated by the minimum of $\{L_R[n]-L_j[n]\}_{n=1}^N$ (Lines \ref{alg1.7}-\ref{alg1.8} in Algorithm 1).

In order to transfer and store the least bytes (energy efficient), a simple neighbor table management is required for the RDC-RMTS (both the root and non-root). When a node has received any time synchronization packets, the node will match the sender address in the neighbor table and allocate memories for the new neighbor. Then node will check if the neighbor is active during each synchronization period by counting the broadcast of the neighbor. Thus the lost neighbor will be delete from neighbor table and only the active neighbor's timestamp will be stored and sent. For the lower message complexity, the buffer message is embedded only in the first broadcast packet of the multi-broadcast.

\emph{\textbf{2) The non-root algorithm:}} The pseudo-code of the non-root algorithm is shown in Algorithm 2.

\LinesNumbered
\begin{algorithm}[ht]
\caption{The pseudo-code for RDC-RMTS. $v_i$ is non-root node, $j\in \mathcal{N}_i, (i,j)\in \mathcal{G}$. Parameter $N$ is the maximal number of multi-broadcast. Parameter $M$ is maximal numbers of neighbor.}
\textbf{Initialization:}                                                                            \\  {\label{alg2.1}}
      \quad{  Set $\hat{\varphi}_{i(MLE)}^j=1$; $\varphi_i=1$}                                      \\  {\label{alg2.2}}
      \qquad{      $\hat{\theta}_{i(MLE)}^j=0$; $\hat{\theta}_i=0$; $n=0$; $ID_i=0$   }             \\  {\label{alg2.3}}
      \qquad{      $\langle\{H_i[n],H_j[n]\}_{n=1}^N \rangle_{old}\leftarrow 0 $        }           \\  {\label{alg2.4}}
      \qquad{      $\langle\{H_i[n],H_j[n]\}_{n=1}^N \rangle_{new}\leftarrow 0 $; $up\leftarrow 0$  }   \\ {\label{alg2.5}}
      \qquad{      $\langle\{L_i[n],L_j[n]\}_{n=1}^N \rangle\leftarrow 0  $; $down\leftarrow 0$     }    \\ {\label{alg2.6}}
      \qquad{      $\{buf_i(m)\}_{m=1}^M\leftarrow 0$                                     }             \\ {\label{alg2.7}}
\BlankLine

\if\textbf{Once $\langle H_j[n], L_j[n], \varphi_j,ID_j,\hat{\theta}_j, buf_j\rangle$ is received:}     \\ {\label{alg2.8}}
{  \eIf{($ID_j>ID_i$)} {                                                                                   {\label{alg2.9}}
        Store $\varphi_j$; $\hat{\theta}_j$; $\langle H_i[n], H_j[n]\rangle_{new}$                      \\ {\label{alg2.10}}
              $up\leftarrow buf_j[i]$; $down\leftarrow \mathop{\min}\limits_{1<n<N}\{L_i[n]-L_j[n]\}$   \\ {\label{alg2.11}}
              $ID_i\leftarrow ID_j$                                                                     \\ {\label{alg2.12}}
        Start parameter estimation compensation task                                                    \\ {\label{alg2.13}}
    }{  Store $buf_i \leftarrow \mathop{\min}\limits_{1<n<N}\{L_i[n]-L_j[n]\}$                          \\  {\label{alg2.14}}
     }
}

\BlankLine
\if\textbf{Upon triggering of compensation task:}{                                                                  \\ {\label{alg2.15}}
    \quad{Calculate $\hat{\varphi}_{i(MLE)}^j$; $\varphi_i\leftarrow\hat{\varphi}_{i(MLE)}^j\times\varphi_j$}       \\ {\label{alg2.16}}
    \quad{Calculate $\hat{\theta}_{i(MLE)}^j$; update $\hat{\theta}_i\leftarrow\hat{\theta}_{i(MLE)}^j+\hat{\theta}_j$ }    \\ {\label{alg2.17}}
    \quad{Logical clock compensation}                                                                                       \\ {\label{alg2.18}}
    \quad{$\langle\{H_i[n], H_j[n]\}_{n=1}^N\rangle_{old}\leftarrow\langle\{H_i[n], H_j[n])\}_{n=1}^N\rangle_{new}$ }       \\ {\label{alg2.19}}
    \eIf {($n<N$) (Rapid flooding)}{                                                                                           {\label{alg2.20}}
        Broadcast $\langle H_i[n], L_i[n], \varphi_i, ID_i, \hat{\theta}_i, buf_i\rangle$                                   \\  {\label{alg2.21}}
        Set $n=n+1$, go back to (if ($n<N$))                                                                                     \\  {\label{alg2.22}}
    }{                                                                                                                          {\label{alg2.23}}
     $n=0$                                                                                                                  \\  {\label{alg2.24}}
     }                                                                                                                        {\label{alg2.25}}
}
\end{algorithm}

\emph{\textbf{3) The rapid-flooding protocol:}} It is employed to minimize the waiting time of the flooding process. Similar as RMTS, $ID_R$ is used to maintain the rapid-flooding protocol. Considering a complex network, there is more than one path between root and another node, then a same time information may be forwarded more than one time, which may results in the communications resources and energy waste.

As discussed above, the longer the flooding waiting time is, the more synchronization errors of multiple network there are due to clock drift. Therefore, it is reasonable that the lower time the flooding path costs, the higher the accuracy of the time information is and the shorter the distances are. The first arrived multi-broadcast packets will be handled by the receiver. As Lines \ref{alg2.8}-\ref{alg2.12} in Algorithm 2, when a node received a group of packets, it will set its local $ID_i$ as the received $ID_j$.

The flooding protocol seems to change complex networks (e.g., grid networks) into a set of simple networks (e.g., Lines or Spanning tree networks). Moreover, the receiver will forward the received time information as soon as possible against the impact of the clock drift and flood waiting time on multi-hop nodes.

\emph{\textbf{4) The relative clock skew estimation:}} The hardware clock timestamps are used to calculate $\hat{\varphi}_{i(MLE)}^j$, i.e. $\langle \{u[n]=H_i[n]-H_j[n]\}_{n=1}^n \rangle_{old}$, $\langle \{v[n]=H_i[n]-H_j[n]\}_{n=1}^n \rangle_{new}$ in (\ref{equ:5}). The details are shown at Line \ref{alg2.19} in Algorithm 2.

According to the discussion in Section II.B, $\varphi_h^{h-1} (1<h\leq k)$ is the relative clock rate between $v_{h-1}$ and $v_h$. The relative clock rate of the $k$ hop node $v_k$ is $\varphi_k^R$, and $\varphi_k^R=\prod_{h=1}^k{\varphi_h^{h-1}}$. Hence, if each node shares its $\hat{\varphi}$ to the neighbors, then the $\hat{\varphi}_k^R$ can be calculated at $v_k$, i.e. $\hat{\varphi}_k^R=\prod_{h=1}^k {\hat{\varphi}_h^{h-1}}$.

Obviously, the $\hat{\varphi}_h^{h-1}$ is exceedingly important to the accurate $\hat{\varphi}_k^R$ over the multiple hop network. In Algorithm 2, the equivalent implementation for the calculation of $\hat{\varphi}_i^R$ is the iteration on $\varphi_i\leftarrow \hat{\varphi}_{i(MLE)}^j\times \varphi_j$ at each hop  (Algorithm 2, line \ref{alg2.16}), i.e. $\hat{\varphi}_i^R$ is $\varphi_i$. If the $\hat{\varphi}_{i(MLE)}^j$ is accurate enough, then RDC-RMTS can establish a consistent clock speed network-wide.

\emph{\textbf{5) The relative clock offset estimation:}} Based on the one-way broadcast, two-way message exchange model and joint skew-offset estimation are used in the RDC-RMTS. In order to structure the two-way message exchange model, the node should collect broadcast messages from all neighbors and store the timestamps. According to the (\ref{equ:12}), only one pair of timestamps are required for the clock offset estimation, thus a node only stores and sends the minimal sample (Lines \ref{alg2.14} in Algorithm 2), i.e., $\mathop{\min}\limits_{1<n<N}\{L_i[n]-L_j[n]\}$.

When a node receives the time information packets, it sets the corresponding sample in the received $buf_j$ as the uplink observation of the two-way message exchange model, and calculates the sample of the corresponding downlink  by the new logical timestamps  (Line \ref{alg2.11}in Algorithm 2), i.e., $\{L_i[n],L_j[n]\}_{n=1}^N$.

Base on the $\hat{\varphi}_{i(MLE)}^j$, and the samples of uplink and downlink, the $\hat{\theta}_{i(MLE)}^j$ is calculated, and is updated to the clock offset compensation parameter $\hat{\theta}_i$ (Line \ref{alg2.17} in Algorithm 2).

\emph{\textbf{6) The logical clock maintenance:}} The RDC-RMTS uses the $\varphi_i$ and $\hat{\theta}_i$ to correct the local logical clock:
$L_i(t+\tau)=L_i(t)+\hat{\theta}_i+(H_i(t+\tau)-H_i(t))\times\varphi_i,\tau>0.$

\emph{\textbf{7) Root election:}} In the flooding synchronization protocols, an specified node (e.g., sink node) is configured as root at the initialization phase. However, when the root is fails, the algorithm cannot synchronize the network unless a new root is elected and creates a reference clock. The simple root election is proposed in the FTSP, and employed in FCSA and RMTS. We refer to the above method as RDC-RMTS, and choose the node which is close to the old root as a new reference.

In conclusion, the central ideas of implementation of RDC-RMTS are:

\emph{i)} multiple one-way broadcast based MLE for clock offset estimation and clock skew estimation, which helps RDC-RMTS against the variable delay and establish more accurate synchronization over large diameter network;

\emph{ii)} clock skew estimation and clock offset estimation rapid sharing, which helps RDC-RMTS to build the same logical clock rate among network, and meets the fast convergence;

\emph{iii)} rapid-flooding, which helps RDC-RMTS against the clock drift and keep the accuracy of shared clock parameters.

\section{Experimental Results and Discussions}
The prototype system is built indoors with 25 Synchronous Sensing Wireless Sensor (SSWS) nodes, which is same as that of RMTS in \cite{RMTS}. The local sync-error, global sync-error, convergence time, and multi-hop sync-error are discussed among the proposed RDC-RMTS and the algorithm in comparison (i.e., FTSP, FCSA, FloodPISync, PulseSync, PulsePISync, and RMTS). It is worth to note that absolute values are used to describe the synchronization error.

\subsection{Testbed Setup}

\textbf{\emph{Testbed overview}}: the \emph{SSWS} node is designed based on the System-on-Chip (SoC) wireless transmission chip CC2530, which uses an enhanced 8051 CPU (8 bits) as the core processor. The clock source frequency of time synchronization algorithm is configured as 1 MHz, which is divided by the system clock source (an external 32 MHz crystal oscillator, the frequency offset may be up to 50 PPM). The MAC Timer2 (which has two portions: a 16 bits timer and a 24 bits overflow counter) of CC2530 is employed to count the clock pulse, and the count value is the hardware clock. The logic clock is defined by the software based on the hardware clock. The timer is configured as up mode with an overflow period of 1 {\textmu s}. The SFD interrupt handling is used to create MAC-Layer timestamp, and the timestamp will be embedded into the MAC payload when a packet is sending.

\textbf{\emph{Sync-error measurement}}: similar to \cite{FTSP,FCSA,PulseSync,RMTS}, a sink-node (connected to a PC) is used to trigger the testbed to create timestamps at the same moment, and it periodically broadcasts test command packets to all of the nodes at fixed intervals. The timestamps are the samples of the synchronization algorithms and are used to calculate the sync-error. Assuming that the wireless signal arrives at all of the test nodes simultaneously, then measurement errors are mainly caused by the differences of the latency that SFD interrupt signal triggers the processor to create timestamps at different nodes. We have extended the above platform and collected a lot of test data to evaluate the measurement error, and the mean and standard deviations of the measurement error are about 0.07 {\textmu s} and 0.0033 respectively. Therefore, the sync-error measurement is credible in our experiments.

\subsection{Experimental Results}
According to the conclusion in the previous flooding time synchronization algorithms, its sync-error mainly depends on the distance of nodes to the reference. Therefore, the experimental results in a large diameter network can fully reflect the performances of a flooding time synchronization algorithm. Considering a complex network (e.g., grid), although there are more than one flooding path to the flooding protocol, but only one of them (the shortest or the fastest) is valid in the flooding time synchronization algorithms. In other words, the flooding time synchronization protocols seem to simplify the complex network into a set of lines. Thus, the evaluation of the flooding time synchronization algorithms is done on a line network.

A line topology network is created in software, i.e., \textbf{R}$\leftarrow$\textbf{1}$\dots\leftarrow$\textbf{24} (diameter of 24), where nodes can only communicate with its neighbors and the sink-node. The specific configurations of the comparison algorithm are as follows.

\emph{i)} The prior delay $\hat{D}_{fixed}$ is 3 {\textmu s} in the FCSA, FloodPISync, PulseSync, PulsePISync, and RMTS.

\emph{ii)} The linear regression table size of the FTSP, FCSA, and PulseSync is eight.

\emph{iii)} The nominal drift of node's clock frequency is $\pm 100$ PPM in the FloodPISync and PulsePISync.

\emph{iv)} The multiple number $N=5$ and the sliding window length $W=2$ (i.e., $\tau=T_b$) in the RMTS and RDC-RMTS.

\emph{v)} The synchronization period (interval) $T_b$ is 30 second and the time interval of error measurement is 10 second.

In order that the test command packet of the sink node reaches all the test nodes almost at the same time and in order to accurately measure the synchronization error, all the nodes must be within each other's communications range and they need to use the same channel and be placed next to each other.  Therefore, the topology of the network is generated by a software and the radio is configured as a slotted CSMA-CA transmit mode to reduce the probability of collision.

If a candidate node fails to receive the time synchronization broadcast for 2 consecutive synchronization rounds, it will begin to apply as a new root immediately, where less than 7 rounds of synchronization periods are required to elect a new root. The synchronization error will increase slightly, but once the new root is elected, the synchronization will recover quickly.

\textbf{\emph{1) Slow-flooding vs. rapid-flooding}}

If node \textbf{1} has been synchronized to \textbf{R}, then the best case is that node \textbf{2} synchronizes to node \textbf{1} at the new broadcast of node \textbf{1}, and node \textbf{2} indirectly synchronizes to \textbf{R}. However, the flooding time cost (or $T_{wait}$) is exceedingly important to the synchronization algorithm to meet such best cast. The time cost may be extremely small (e.g. $T_{wait}\rightarrow0$), or approximately equal to a synchronization interval ( e.g. $T_{wait}\rightarrow T_b$).
In the slow-flooding approaches, any nodes in the line are periodically and asynchronously triggered to broadcast time information. Then, the best case for slow-flooding is $T_{wait}\rightarrow0$, and the worst case is $T_{wait}\rightarrow T_b$. A time information of root may cost up to $24\times T_b $ to forward to the multiple-hop node \textbf{24}. However, in the rapid-flooding model, the time information of \textbf{R} can be quickly forwarded to node \textbf{24} in the time of less than $T_b$.

\textbf{\emph{2) Local synchronization error}}

\begin{figure}[!htb]
\centering
\includegraphics[scale=0.6]{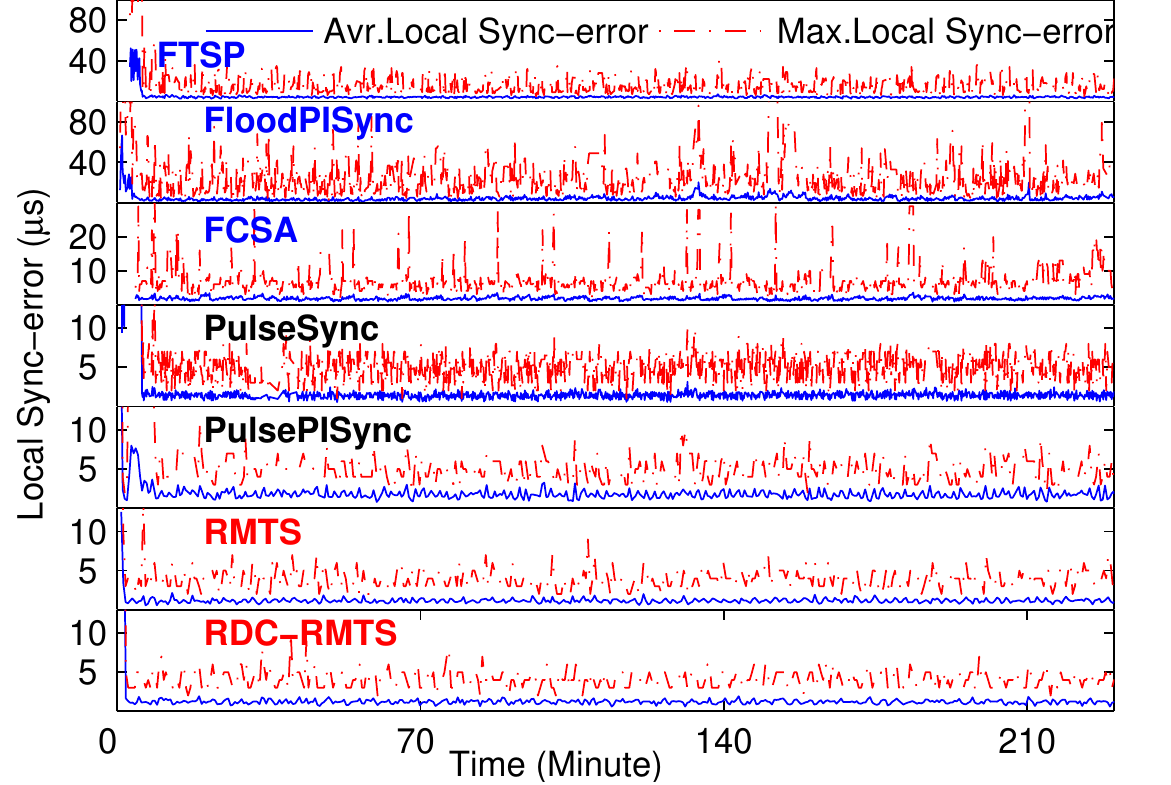}
\caption{Maximal local synchronization error in line of the FTSP, FCSA, PulseSync, FloodPISsync, PulsePISync, RMTS, and RDC-RMTS.}
\label{LocalSynchError}
\end{figure}

\begin{figure}[!htb]
\centering
\includegraphics[scale=0.6]{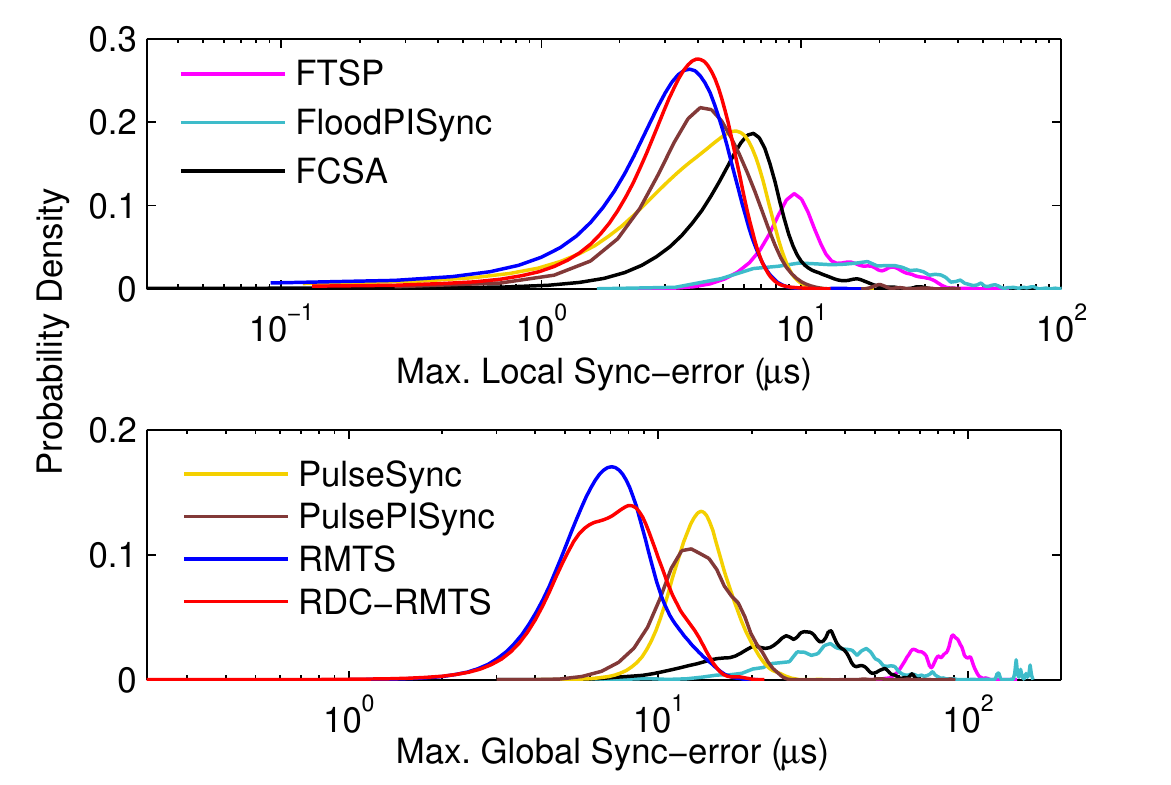}
\caption{The probability density of maximal local synchronization error and maximal global synchronization error. Logarithmic coordinates are used for X-axis to show more details.}
\label{Local_global_PDF}
\end{figure}

\begin{figure}[!htb]
\centering
\includegraphics[scale=0.7]{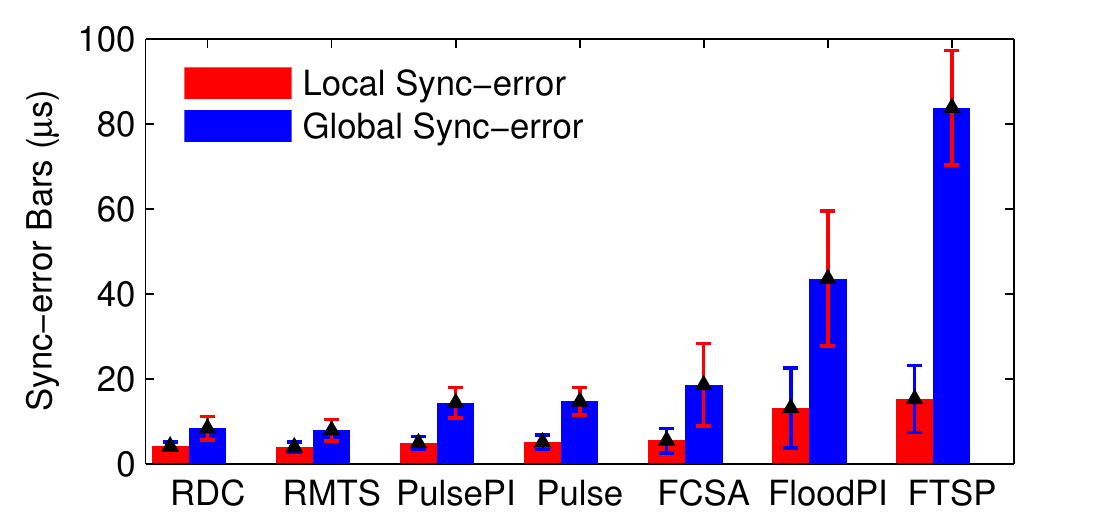}
\caption{Error bars of maximal local synchronization error and maximal global synchronization error.}
\label{Local_Global_Errorbar}
\end{figure}

A local sync-error is calculated based on the measurement timestamps of pairs of adjacent nodes. Therefore, it is used to describe the synchronization error between any adjacent nodes. Moreover, it can be used to analyze performance of the clock offset estimation and clock skew estimation. As shown in Fig. \ref{LocalSynchError}, the experimental results indicate the instantaneous average and instantaneous maximum of local synchronization error. The probability density of the maximal local synchronization error is shown in upper panel of Fig. \ref{Local_global_PDF}. The time-average and standard deviation of maximal local sync-error are shown as the red bars in Fig. \ref{Local_Global_Errorbar}.
When time synchronization algorithm has finished converging, e.g., after 30  minutes, we calculate all of the statistical characteristics.

The local synchronization errors of slow-flooding approaches (FTSP, FCSA, and FloodPISync) are larger than that of the rapid-flooding time synchronization algorithms. As discussed in Section III.D, the waiting time of the slow-flooding results in the inaccurate time synchronization information and leads more synchronization error to the neighboring nodes. According to the summary in Table \ref{tab:1} and considering the FCSA and PulseSync, the difference between them is the used of flood protocol (slow-flooding, or rapid-flooding). However, the local synchronization error in PulseSync is less than that in FCSA. In PulseSync, the probability density of the local synchronization error is closer to zero, and the mean and standard deviation are smaller.

Considering the rapid-flooding approaches, both the clock skew estimation and clock offset estimation are different from each other (as shown in Table \ref{tab:1}). The RMTS and RDC-RMTS have lower and smoother instantaneous local synchronization error, and their maximal local synchronization error probability density is tighter and closer to zero. The 95\% confidence interval of mean and standard deviation for maximal local synchronization error are 4.7-4.96 {\textmu s} and 1.433-1.619 in PulsePISync, 4.97-5.15 {\textmu s} and 1.667-1.798 in PulseSync, respectively, which are 4.05-4.25 {\textmu s} and  0.983-1.095 in RDC-RMTS, 3.82-4.02 {\textmu s} and  1.08-1.215 in RMTS.

It is clear that, the clock skew MLE in (\ref{equ:6}) results in better performance on local synchronization of the RDC-RMTS and RMTS, and the joint clock skew-offset MLE in (\ref{equ:12}) leads the local synchronization of RDC-RMTS to be smoother.

\vspace{0.3cm}

\vspace{0.3cm}

\textbf{\emph{3) Global synchronization error}}

\begin{figure}[!htb]
\centering
\includegraphics[scale=0.6]{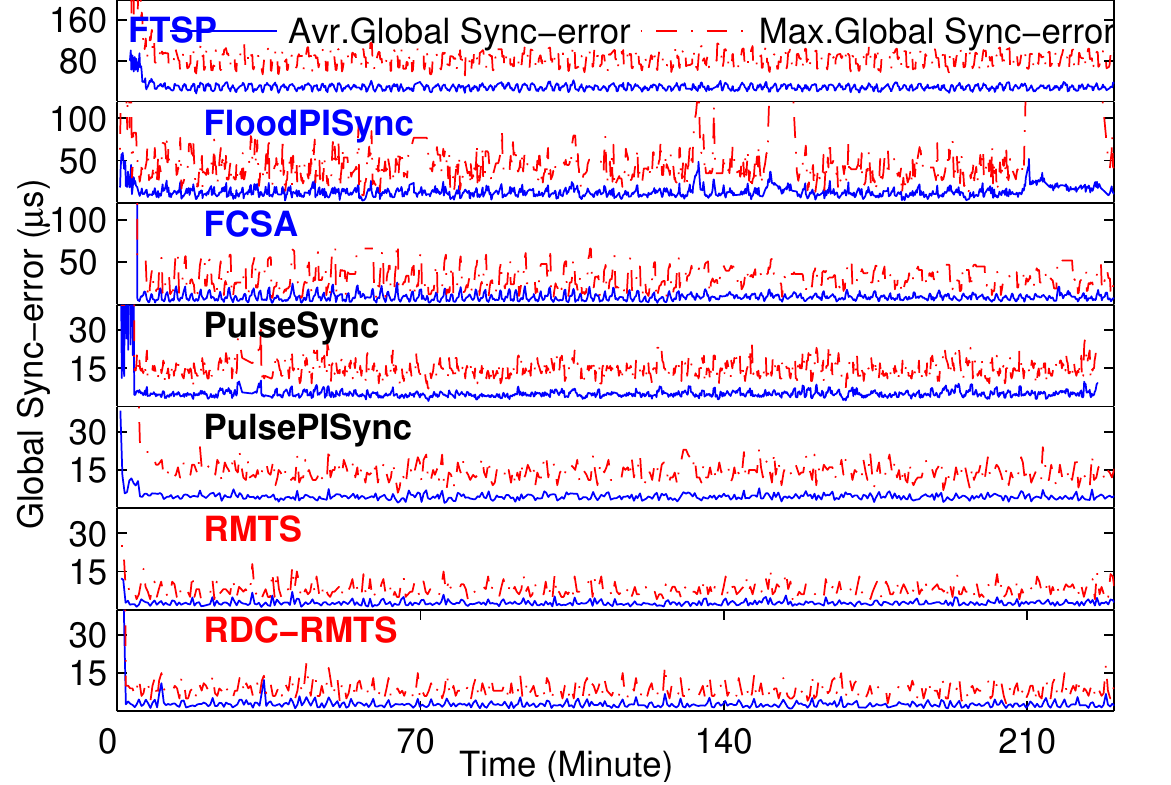}
\caption{Maximal global synchronization error in line of the FTSP, FCSA, PulseSync, FloodPISsync, PulsePISync, RMTS, and RDC-RMTS.}
\label{GlobalSynchError}
\end{figure}

\begin{figure}[!htb]
\centering
\includegraphics[scale=0.6]{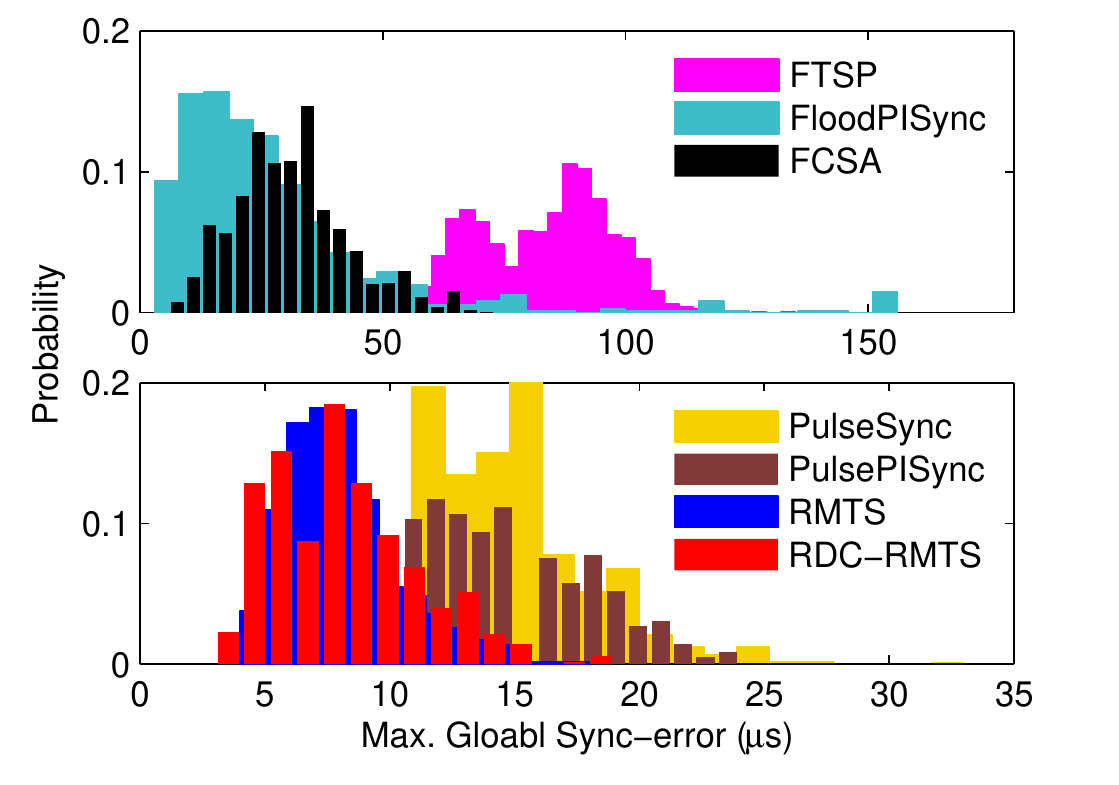}
\caption{The actual distributions of maximal global synchronization error.}
\label{GlobalHIST}
\end{figure}

A global synchronization error is calculated based on the measurement timestamps of an arbitrary pair of nodes. Therefore, its instantaneous value is used to describe the synchronization error of the network. As shown in Fig. \ref{GlobalSynchError}, the experimental results indicate the instantaneous average and instantaneous maximum of global synchronization error. The probability density of the maximal local sync-error is shown in lower panel of Fig. \ref{Local_global_PDF}. Considering the maximal local synchronization error, the time-average and standard deviation are shown as the blue bars in Fig. \ref{Local_Global_Errorbar}, and the actual distributions are shown in  Fig. \ref{GlobalHIST}.

The RMTS and RDC-RMTS have lower and smoother instantaneous global synchronization error, and their maximal global synchronization error probability density is tighter and closer to zero. The 95\% confidence interval of mean and standard deviation for maximal global synchronization error are 14.04-14.65 {\textmu s} and 3.329-3.756 in PulsePISync, 14.52-14.88 {\textmu s} and 3.103-3.357 in PulseSync, respectively, which are 8.14-8.55 {\textmu s} and  2.579-2.870 in RDC-RMTS, 7.70-8.11 {\textmu s} and  2.344-2.636 in RMTS.

It is clear that, the instantaneous global synchronization error in RDC-RMTS is almost the same as that in RMTS, but much lower than those in the remaining algorithms. According to the experimental results (both of the local and global synchronization error) and the summary in Table \ref{tab:1}, it can be summarized that:

\emph{i)} the rapid-flooding time synchronization algorithm is better than the slow-flooding one;

\emph{ii)} the clock skew MLE in (\ref{equ:6}) is more accurate than the linear regression in PulseSync and the PI clock skew estimation in PulsePISync;

\emph{iii)} the proposed joint clock skew-offset MLE in RDC-RMTS is as accurate as the fixed $\hat{D}_{fixed}$ clock offset estimation compensation in RMTS.

\vspace{0.3cm}

\vspace{0.3cm}

\vspace{0.3cm}

\textbf{\emph{4) Convergence time}}

The convergence time is important to the time synchronization of a large-scale network. It mainly depends on the network diameter in flooding time synchronization algorithm. In other words, the larger the network diameter is, the longer the convergence time is. Moreover, the flooding protocol and clock skew estimation will impact the convergence of the algorithm.

In our experiments, we assume that the topology of the network and reference node are fixed, then we randomly power on the node within 30 seconds. More than 50 round experimental results are used to calculate the convergence time. The experimental results are shown in Table \ref{tab:3}.

The accurate clock skew  estimation can be generated at the second multiple-one-way-broadcast process by using MLE in (\ref{equ:6}), meanwhile the rapid-flooding protocol may lead the RMTS and RDC-RMTS to converge immediately, i.e., converging in the second sync-period. Benefiting from the clock skew estimation and clock offset estimation rapid sharing mechanism,  RDC-RMTS can converge with great probability during the third synchronization period.

\vspace{0.8cm}
\begin{table}[htbp]\scriptsize
 \centering
 \captionsetup{justification=centering}
 \caption{\\Summary of the experimental results. The convergence time of each algorithm is calculated from more than 50 round experiments. The maximal global synchronization error after the synchronization algorithm converges is indicated below.}{\label{tab:3}}
 {
 \begin{tabular}{ccccccc}
  \toprule
  \toprule
              & \textbf{Convergence Time}        & \multicolumn{3}{c} { \textbf{Max. Global Sync-error (\textmu s)}}          \\
              & {(Sync-period)}                       &{ Minimal}         &{ Mean}          &{Maximal}\\
  \midrule
FTSP	       &$>$12	                         &51	 	         &83               &116	     \\
FCSA	       &$>$12	                         &6	   	             &31               &73	     \\
FloodPISync    &8-11                             &9	   	             &55               &159	      \\
PulseSync      &8-10                             &7	   	             &15               &33	      \\
PulsePISync    &8-10                             &6	   	             &14               &24	      \\
RMTS           &2-6	                             &4	   	             &8                &18	      \\
RDC-RMTS       &2-7	                             &3	   	             &8                &19	      \\
  \bottomrule
  \bottomrule
 \end{tabular}
 }
\end{table}
\vspace{0cm}

\textbf{\emph{5) Synchronization error to root}}

\begin{figure}[!htb]
\centering
\includegraphics[scale=0.6]{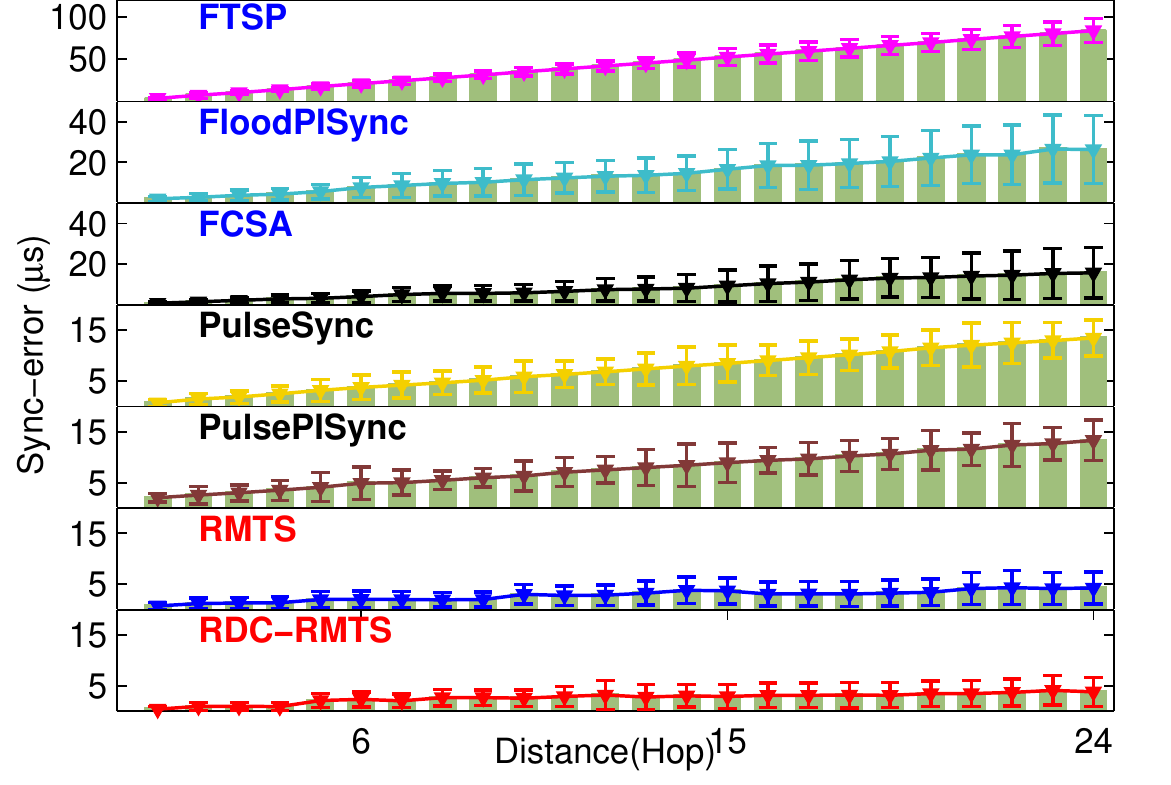}
\caption{Time maximal sync-errors from reference are described with error bar plot. Bars indicate standard deviation over time.}
\label{MultiHopErrorBar}
\end{figure}

As discussed above, the diameter of the network has a great influence on the accuracy of the flooding time synchronization algorithm. The synchronization error to root commonly increases as the growth of diameter, such as the exponential growth in FTSP, and square root growth in PulseSync, PISync and FCSA. Obviously, the growth speed of the synchronization error is slower, then the flooding time synchronization algorithm may achieve better performance in a large-scale WSN. The experimental results are shown in Fig. \ref{MultiHopErrorBar}, in which the time-average of maximal synchronization error to the reference are reported. As the error analysis in Section III.D, the RDC-RMTS and RMTS have significantly reduced the possible by-hop error accumulation, and the growth of their synchronization errors is significantly slower than other algorithms.

\vspace{0.3cm}

\subsection{Synchronization Error vs. Sync-period}
The sync-period of time synchronization algorithm is as important as clock skew estimation and clock offset estimation. Assuming that clock skew is fixed, the larger the sync-period is, the more accurate the clock skew estimation is and the lower the energy consumption is. However, the clock drift is always changing due to the changes in environment (e.g., temperature or voltage), thus the assumption that the relative clock is fixed is not always true. In other words, the time synchronization algorithm should re-synchronize the network at appropriate intervals. Clearly, the frequent re-synchronization will lead to high hardware resources cost, e.g. communications resource, and energy. Considering the energy-constrained WSNs, balancing energy consumption and synchronization accuracy is important to achieve a long life.

\begin{figure}[!htb]
\centering
\includegraphics[scale=0.6]{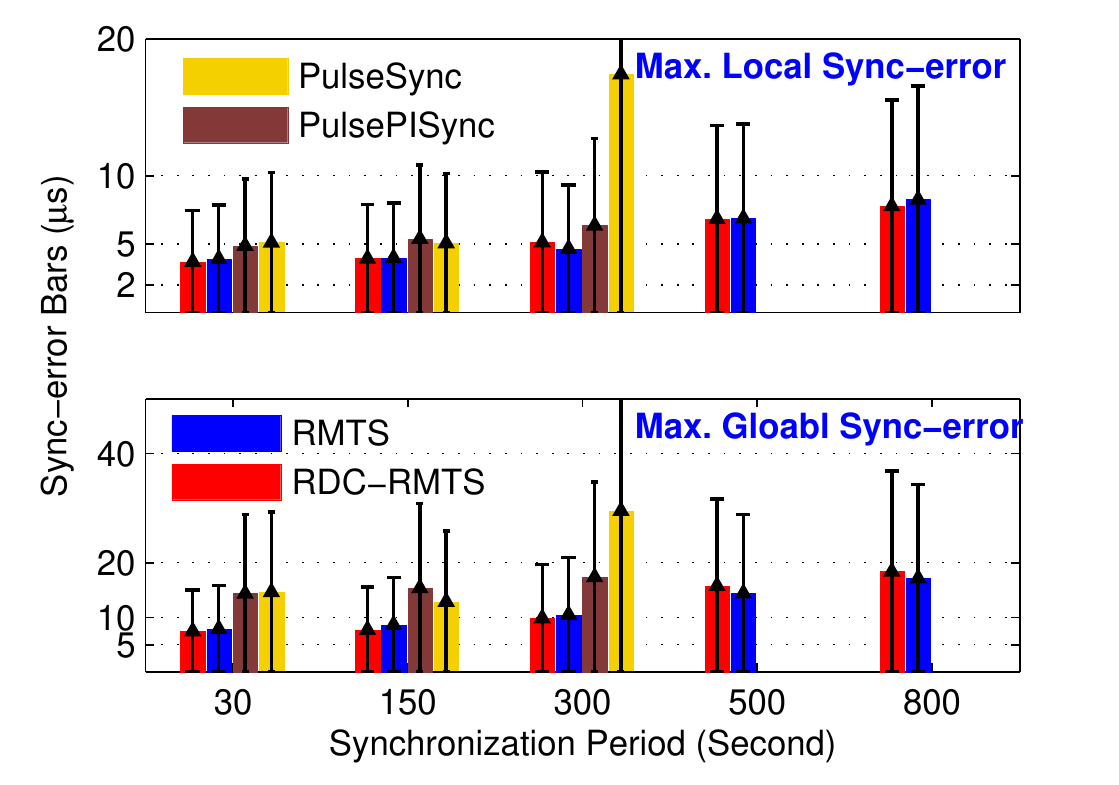}
\caption{Time average of the absolute maximal local sync-error and absolute maximal global sync-error at different synchronization interval.}
\label{Mul_Local_Global_ErrorBar}
\end{figure}

Therefore, the time synchronization algorithm using less broadcasts to build more accuracy is more energy-efficient. In this part, we evaluate the RDC-RMTS at different sync-periods, i.e., 30 seconds, 150 seconds, 300 seconds, 500 seconds, and 800 seconds. In other words, the algorithm is evaluated at different energy efficiencies. According to the experimental results discussed above, we only discuss the rapid-flooding based approaches. Figure \ref{Mul_Local_Global_ErrorBar} shows the mean and standard deviation of the time-average of maximal global synchronization error.

According to the experimental results, it is clear that the RDC-RMTS is more accurate than PulseSync and PulsePISync when they use the same synchronization period. Moreover, when they use the same number of broadcasts (consistent energy efficiency), the RDC-RMTS is still more accurate, e.g., for instance, the accuracy of the 30 seconds sync-period in  PulseSync and PulsePISync equals to that of 150 seconds sync-period in RDC- RMTS and RMTS. Even with a sync-period of 500 seconds, the accuracy in RDC-RMTS is almost the same as that of the PulseSync and PulsePISync with a sync-period of 30 seconds.

\vspace{0.3cm}

According to the above experimental results and discussions, we can briefly summarize the improvements of the proposed RDC-RMTS as follows:

\emph{i)} using the proposed joint clock skew-offset MLE to obtain the real-time $\hat{D}_{fixed}$, the RDC-RMTS is accurate, and it is better scalability than the approaches using fixed $\hat{D}_{fixed}$ (i.e., $\bar{D}$ );

\emph{ii)} the RDC-RMTS is a rapid-flooding time synchronization protocol, meanwhile, the clock skew MLE and parameters sharing mechanism guarantee fast convergence of network-wide and slow growth of by-hop error accumulation on multi-hop nodes.

{

\section{Simulations in Complex Networks}
The basic network unit in the flooding time synchronization approaches is line, and it is expected that the flooding protocol always simplifies a complex network to lines. Ideally, it is assumed that the communications is reliable, and the flooding time synchronization algorithms are most concerned with the by-hop error accumulation problem, which is determined by the diameter of the network, the distance between the reference and the multi-hop nodes, and flooding latency. This is why lines are popularly used to evaluate the performances in the existing flooding time synchronization researches.

However, the structure and density of the networks indirectly affect the performances of flooding time synchronization algorithms. Considering an actual WSN, it is an unreliable wireless network, in which the communications collision probability depends on the density of network and actual payload of the channel, e.g., the calculation of send success probability in pure ALOHA protocol. Specifically, the communications collision probability in complex network is relative higher than the line due to the higher density.

Therefore, the actual performances of flooding time synchronization algorithms should to be further discussed in complex networks. However, considering a large-scale WSN with randomly deployed topology, it is almost impossible to evaluate time synchronization algorithm by using testbed. In most of the existing researches, they deploy all of the nodes in a small area to ensure that they can receive the same synchronization error measurement message at the same moment, and model the topology in a software. Therefore the additional interference between nodes (non-adjacent nodes in the topology) is serious and will result in incorrect experimental results. Obviously, one can only use lightweight grid testbed (e.g., $5\times5$, $5\times7$) to model the complex network. e.g., FTSP, PulseSync, FCSA, RMTS, ATS, and MTS.

In view of the above discussions, the simulations in large-scale WSN with randomly-deployed topology are employed to evaluate the flooding time synchronization algorithms. In the following part, the results of flooding protocol (rapid-flooding and slow-flooding) in the reliable and unreliable communications WSNs are discussed, separately, and the actual distance of the flooding path is discussed in detail.

A strongly connected network with 300 nodes is built, where the communications range of nodes is 80 meter, and nodes are randomly deployed in a rectangular area of $200\times200$ square meters. More than 10,000 observations of runs are collected for each simulation. The synchronization interval is 30 seconds. The flooding latencies at each node are random variables, which are among (0.01, 0.05) seconds in rapid-flooding, and (0.01, 30) seconds in slow-flooding. Hop number of the shortest path between the reference and a node is the node's distance, then the diameter of the network is the longest distance.

Figure \ref{FloodingPathGenration} shows one of the flooding path structures of the referenced time information in the randomly-deployed network, which is a spanning tree. For the same referenced time information, one can find only one line (valid flooding path) between the reference and nodes. Considering the complete simulation, the line of a node may be different from others. Thus, the by-hop error accumulation problem, the performances of the algorithm highly depend on the longest line. Moreover, the clock skew estimation should be recalculated when a different path is selected by algorithm, and the observation set for clock skew estimation need to re-load. During these periods, the global synchronization error may increase due to the clock drift. Therefore, the stable synchronization algorithms require the clock skew estimation to converge fast. To the linear regression used in FTSP, FCSA, and PulseSync, the new clock skew estimation converges after a number of synchronization periods (depending on size of the regression table, e.g., 8), while the clock skew MLE algorithm in RDC-RMTS may converge during the second synchronization period \cite{MLE}.

\begin{figure}[!htb]
\centering
\includegraphics[scale=0.6]{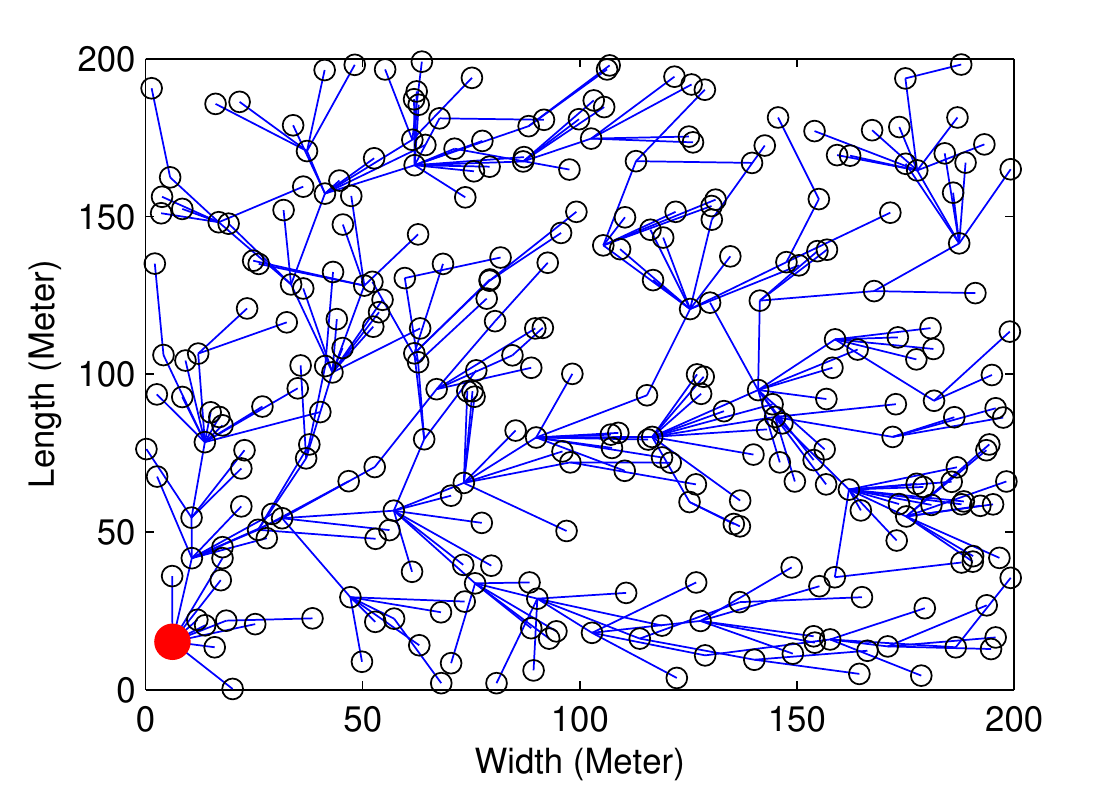}
\caption{A demonstration of the flooding protocol in the randomly-deployed network (300 nodes). The red dot is the reference, and the diameter of the networks is 9.}
\label{FloodingPathGenration}
\end{figure}
\vspace{0.3cm}

\begin{figure}[!htb]
\centering
\includegraphics[scale=0.6]{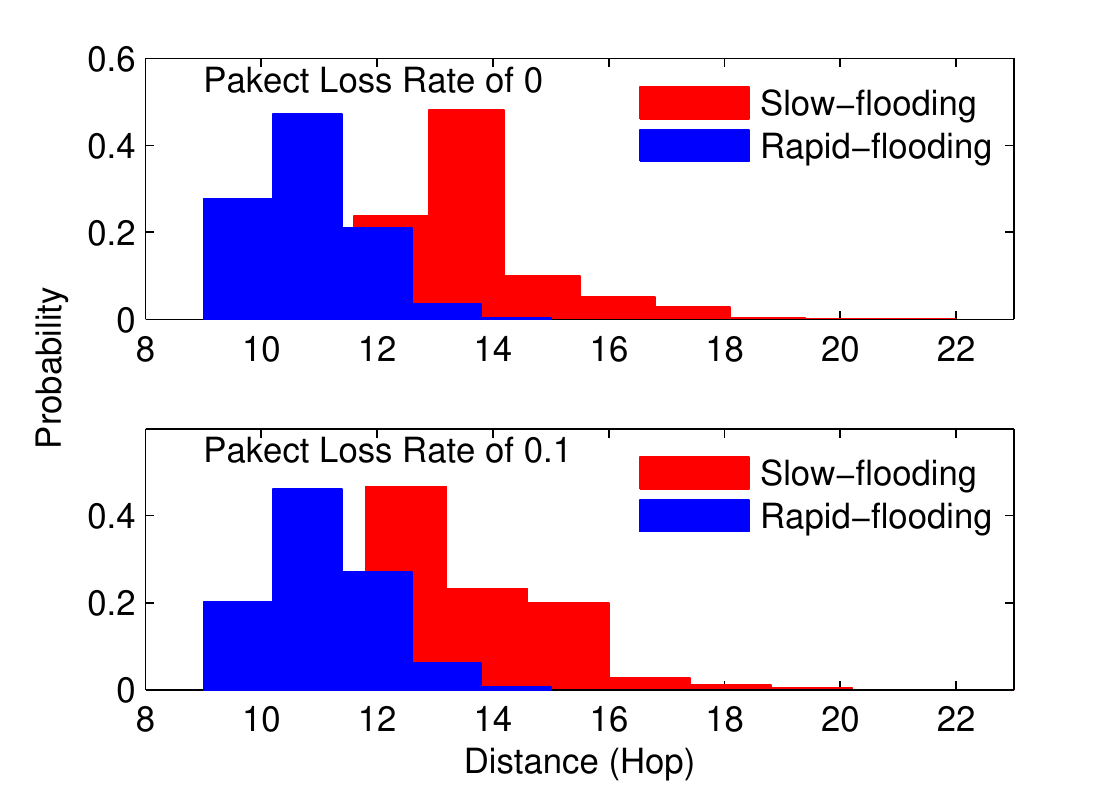}
\caption{The distribution of the longest flooding path in the reliable randomly-deployed network (diameter of 9).}
\label{DistanceIeal}
\end{figure}
\vspace{0.3cm}

The distance of the longest flooding path is calculated for every synchronization round. The simulations of the reliable randomly-deployed network are shown in upper panel Fig. \ref{DistanceIeal}, where the maximal and minimal distances in rapid-flooding are 15 and 9 hops, respectively, and in slow-flooding they are 22 and 9 hops. The results of the unreliable randomly-deployed network (packet loss rate (PLR) of 0.1) are shown in lower panel Fig. \ref{DistanceIeal}, where the maximal distances in slow-flooding increase to 23. The summary of the results are shown in Table \ref{tab:4}. Obviously, the probability of the shortest flooding path (distance of 9) is much higher in rapid-flooding.

To the large-scale complex network, the following conclusions are drawn,

\emph{i)} whether using rapid-flooding or slow-flooding, the actual flooding-path is longer than the diameter of the network;

\emph{ii)} the actual flooding-path will become longer results from the increases in unreliable communications connections;

\emph{iii)} the rapid-flooding is more possible to carry out the shorter flooding path than slow-flooding.

In a way of increasing the flooding path, the large-scale complex network poses challenge to the flooding time synchronization algorithms. As a result, the by-hop error accumulation problem will become to be more serious and outstanding. Aimed at alleviating this problem, the RDC-RMTS focuses on minimizing the multi-hop synchronization error caused by delay and clock drift, and uses accurate clock parameters estimation to restrain amplification of synchronization error caused by increase in flooding-path. The experimental results indicate that the RDC-RMTS is still accurate in a large diameter line.

\vspace{0.8cm}
\begin{table}[htbp]\scriptsize
 \centering
 \captionsetup{justification=centering}
 \caption{\\The probability of flooding path distance (partial). The maximal probability of distances are at 13 hops and 11 hops in slow-flooding and rapid-flooding, respectively.}{\label{tab:4}}
 {
 \begin{tabular}{ccccccccc}
  \toprule
  \toprule
                    & \multicolumn{4}{c}{ \textbf{Probability }}                                                          \\
  \textbf{Distance} &  \multicolumn{2}{c}{ \textbf{Slow-flooding }}       &  \multicolumn{2}{c}{ \textbf{Rapid-flooding }} \\
       (Hop)        &{ PLR of 0}           & { PLR of 0.1}           &{ PLR of 0}                & { PLR of 0.1}      \\
  \midrule
    9	            & $<$0.00002           &$<$0.000009	 	         & 0.02875                    & 0.015655    \\
    11	            & 0.092142	           &0.053217	  	         & 0.47318                    & 0.46054    \\
    13              & 0.28063              &0.27917	                 & 0.035833                   & 0.060601    \\
    15              & 0.098958             &0.13517	   	             & $<$0.00009                 & 0.000334    \\
    17              & 0.019475             &0.027861	  	         & 0                          & 0    \\
    19              & 0.003042             &0.003791	  	         & 0                          & 0    \\
  \bottomrule
  \bottomrule
 \end{tabular}
 }
\end{table}

}

\section{Conclusions }
In this paper, the rapid-flooding multi-broadcast time synchronization algorithm with real-time delay compensation (RDC-RMTS) is proposed. The synchronization error in flooding time synchronization is analyzed in detail, and the by-hop error accumulation problem, which is determined by distance of flooding path, flooding latency, packet delay, and clock drift, is accused of being the major challenges. Moreover, the main disadvantages of flooding time synchronization in a large-scale complex network are summarized, i.e., increases in distance of flooding path. In the light of this, the RDC-RMTS employs rapid-flooding and accurate clock parameter estimations against the error accumulation on the multi-hop, where the synchronization error due to delay and clock drift is minimized significantly.

The significant difference between the RDC-RMTS and the existing flooding time synchronization algorithms is that, a joint clock skew-offset estimation is used to remove delay from clock offset estimation; and an improved two-way message exchange model is constructed in the one-way broadcast-based flooding synchronization. The possible collision due to increase in message complexity of RDC-RMTS is carefully considered, and an innovative implementation is developed to guarantee the reliability of RDC-RMTS, where it can subtly compensates the clock offset estimation by using the existing delay estimates. The RDC-RMTS is compared to the existing flood time synchronization algorithms in a variety of ways. The experimental results show that the local synchronization error, global synchronization error, convergence time, and hop-by-hop error accumulation are significantly improved. Moreover, by using the real-time delay compensation, the RDC-RMTS may better meet the accurate time synchronization requirements in the large-scale WSN.

\ifCLASSOPTIONcaptionsoff
  \newpage
\fi
{
\scriptsize
\bibliographystyle{ieeetr}
\bibliography{CYB}
}

\begin{IEEEbiography}[{\includegraphics[width=1in,height=1.25in,clip,keepaspectratio]{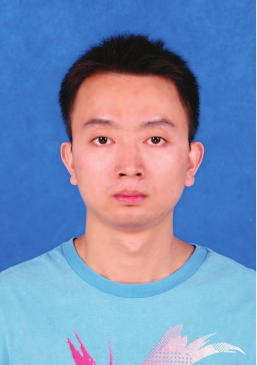}}]{Fanrong Shi}
(S'18, M'20) has graduated from Southwest University of Science and Technology (SWUST), Mianyang, China, having received his B.E. degree in Communications Engineering in 2009, M.E. degree in Communications and Information System in 2012, and Ph.D. degree in Control Science and Engineering in 2019 respectively. Currently He is a member of Robot Technology Used for Special Environment Key Laboratory of Sichuan Province, and an Assistant Professor with the School of Information Engineering at SWUST. He has been dedicating himself to the research of time synchronization and location in wireless sensor networks applications, internet of things, industrial wireless network, wireless synchronous measurement and acquisition, and intelligent instruments.
\end{IEEEbiography}

\begin{IEEEbiography}[{\includegraphics[width=1in,height=1.25in,clip,keepaspectratio]{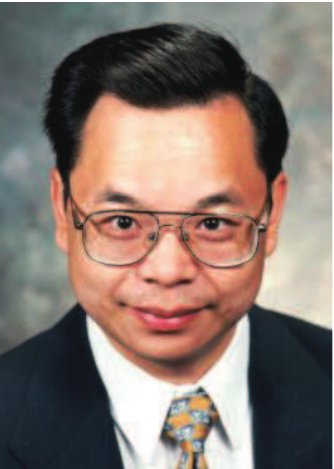}}]{Simon X. Yang}
(S'97, M'99, SM'08) received the B.Sc. degree in engineering physics from Beijing University, China in 1987, the first of two M.Sc.  degrees in biophysics from Chinese Academy of Sciences, Beijing, China in 1990, the second M.Sc. degree in electrical engineering from the University of Houston, USA in 1996, and the Ph.D. degree in electrical and computer engineering from the University of Alberta, Canada in 1999. Currently he is a Professor and the Head of the Advanced Robotics and Intelligent Systems (ARIS) Laboratory at the University of Guelph in Canada. His research interests include intelligent systems, robotics, sensors and multi-sensor fusion, wireless sensor networks, control systems, and computational neuroscience. Prof. Yang serves as the Editor-in-Chief of International Journal of Robotics and Automation, and Associate Editor of IEEE Transactions on Cybernetics, IEEE Transactions on Artificial Intelligence, and several other journals. He has involved in the organization of many conferences.
\end{IEEEbiography}

\begin{IEEEbiography}[{\includegraphics[width=1in,height=1.25in,clip,keepaspectratio]{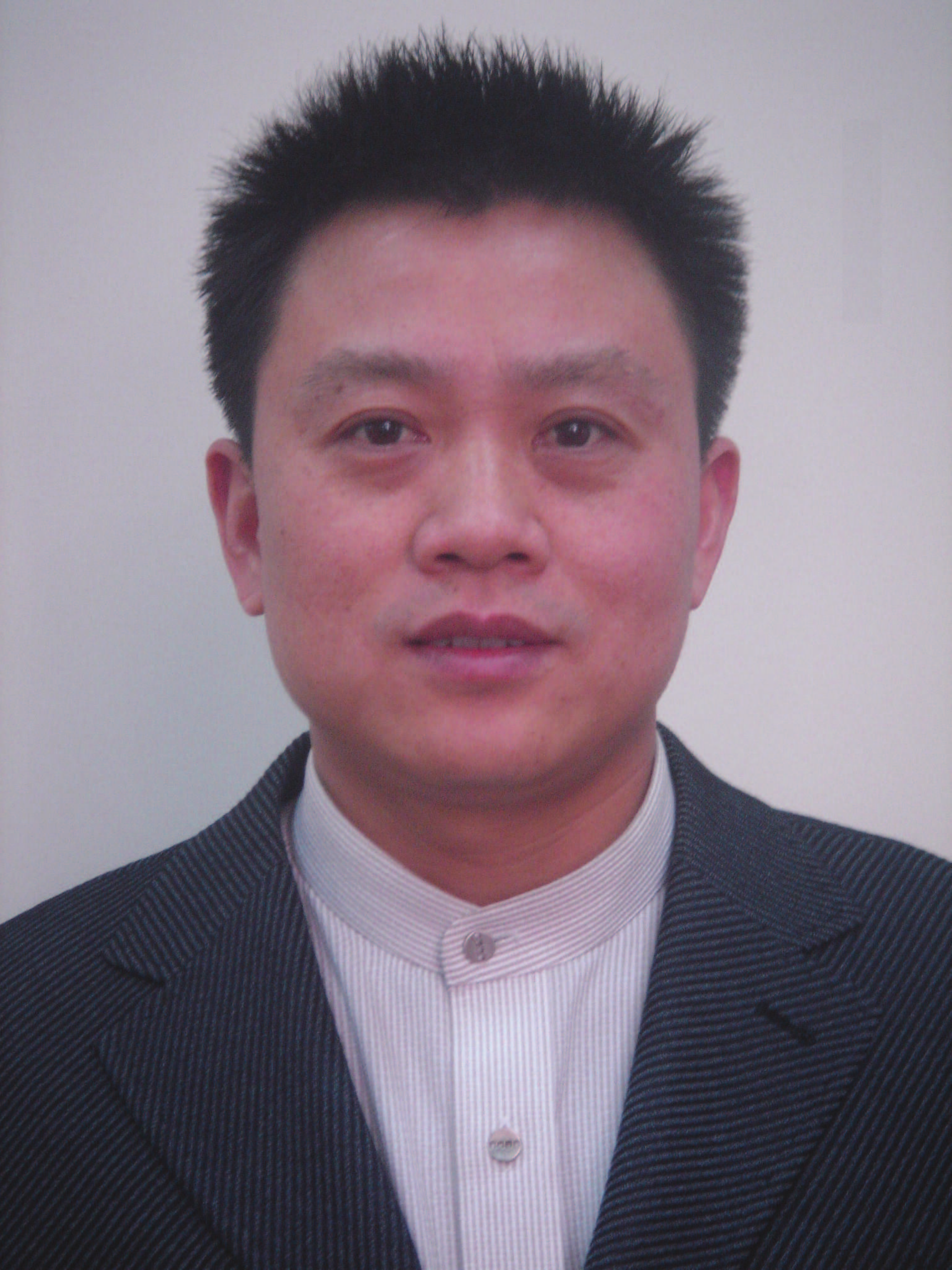}}]{Xianguo Tuo}
received the B.E., M.E. and Ph.D. degrees in Nuclear Geophysics, Environmental Radiation Protection, Applied Nuclear Technology in Geophysics from the Chengdu University of Technology in Chengdu, China, in 1988, 1993, 2001 respectively. From 2006 to 2007, he worked as visiting scholar at the School of Bioscience at University of Nottingham, UK. Since 2001, he is Professor with College of Nuclear Technology and Automation Engineering, Chengdu University of Technology, China. Since 2012, he becomes Professor with the School of National Defense Science and Technology, Southwest University of Science and Technology in Mianyang, Sichuan, China. Currently, he is the Professor and President of Sichuan University of  Science and Engineering, Zigong, Sichuan. Professor Tuo received The National Science Fund for Distinguished Young Scholars in 2011. Currently, his research interests are in detection of radiation, seismic exploration and specialized robots.
\end{IEEEbiography}

\begin{IEEEbiography}[{\includegraphics[width=1in,height=1.25in,clip,keepaspectratio]{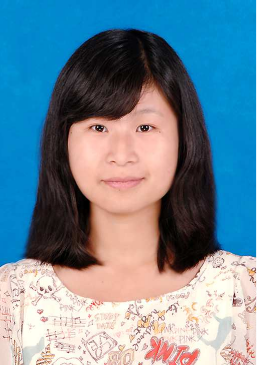}}]{Lili Ran}
received B.E. in Electric Engineering, in 2010 , and M.E. in Rail Transit and Electrical Automation from the Southwest Jiaotong University, Chengdu, China, in 2013. She is currently a member of the Robot Technology Used for Special Environment Key Laboratory of Sichuan Province, and a lecturer in control science and engineering with the Department of School of Information Engineering at Southwest University of Science and Technology. Her research interest is numerical simulation distributed sensing and system modelling.
\end{IEEEbiography}

\begin{IEEEbiography}[{\includegraphics[width=1in,height=1.25in,clip,keepaspectratio]{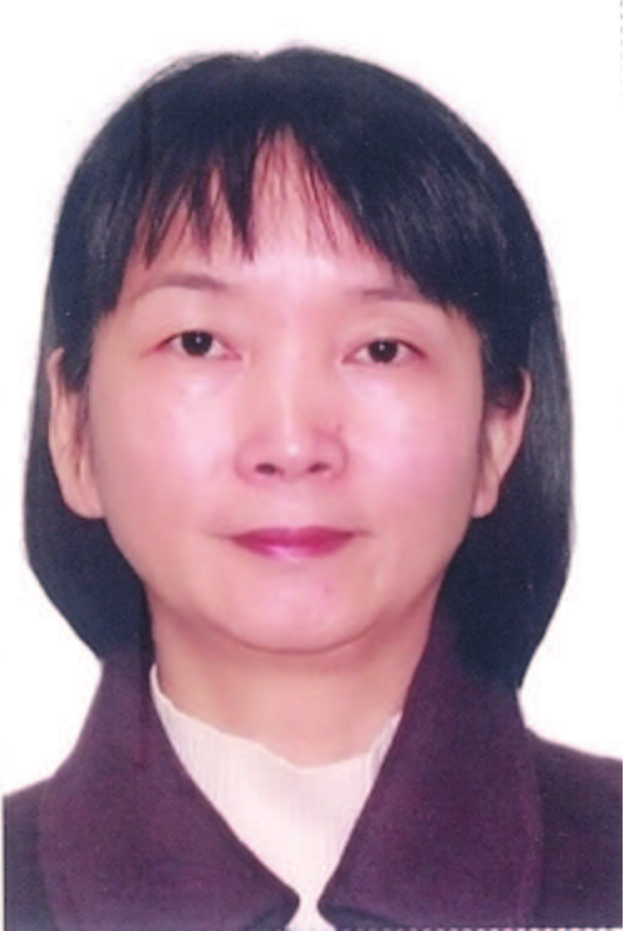}}]{Yuqing Huang}
received M.E. in Industrial Automation from Sichuan University, Chengdu, China, in 1994. She is currently a member of the Robot Technology Used for Special Environment Key Laboratory of Sichuan Province, and a Professor in Control Science and Engineering with the Department of School of Information Engineering at Southwest University of Science and Technology, Mianyang, China. Her research interests include intelligent information processing, machine vision processing, and robot technology.
\end{IEEEbiography}

\end{document}